# Tuning Structural Modulation and Magnetic Properties in Metal-Organic Coordination Polymers $[CH_3NH_3]Co_xNi_{1-x}(HCOO)_3$

Authors

**Madeleine Geers[ab], Oscar Fabelo[a], Matthew J. Cliffe[b] and Laura Cañadillas-Delgado[a]***

[a]Diffraction Group, Institut Laue Langevin, 71, avenue des Martyrs, Grenoble, 38042, France

[b]School of Chemistry, University Park, Nottingham, NG7 2RD, UK

Correspondence email: lcd@ill.fr

**Funding information** Institut Laue Langevin (studentship No. PhD-201_26 to Madeleine Geers); School of Chemistry, University of Nottingham (bursary No. Hobday bequest to Matthew J. Cliffe).

**Synopsis** We show that the modulated phase transitions in the solid solutions $[CH_3NH_3]Co_xNi_{1-x}(HCOO)_3$, with x = 0.25 (**1**), 0.50 (**2**) and 0.75 (**3**), can be tuned by the metal ratio, using variable temperature single-crystal and powder neutron diffraction measurements and bulk magnetometry.

**Abstract** Three solid solutions of $[CH_3NH_3]Co_xNi_{1-x}(HCOO)_3$, with $x$ = 0.25 (**1**), 0.50 (**2**) and 0.75 (**3**), were synthesized and their nuclear structures and magnetic properties were characterized using single crystal neutron diffraction and magnetization measurements. At room temperature, all three compounds crystallize in the *Pnma* orthorhombic space group, akin to the cobalt and nickel end series members. Upon cooling, each compound undergoes distinct series of structural transitions to modulated structures. Compound **1** exhibits a phase transition to a modulated structure analogous to the pure nickel compound (Canadillas-Delgado *et al.*, 2020), while compound **3** maintains the behaviour observed in the pure cobalt compound reported by Canadillas-Delgado *et al.*, 2019, although in both cases the temperatures at which the phase transitions occur differ slightly from the pure phases. Monochromatic neutron diffraction measurements showed that the structural evolution of **2** diverges from that of either parent compound, with the competing hydrogen bond interactions which drive the modulation throughout the series producing a unique sequence of phases. It involves two modulated phases below 96(3) K and 59(3) K, with different **q** vectors, similar to the pure cobalt compound (with modulated phases below 128 K and 96 K), however it maintains the modulated phase below magnetic order (at 22.5(7) K), resembling the pure nickel compound (which present magnetic order below 34 K), resulting in an improper modulated magnetic structure. Despite these large scale structural changes, magnetometry data reveal that the bulk magnetic properties of these solid solutions form a linear continuum between the end members. Notably, doping of the metal site in these solid solutions allows for tuning of bulk magnetic properties, including magnetic ordering temperature, transition temperatures, and the nature of nuclear phase transitions, through adjustment of metal ratios.

## 1. Introduction

Coordination polymers (CPs) can integrate multiple physical properties into a single framework (Cui *et al*., 2016; Lin *et al*., 2014; Li *et al*., 2016; Liu *et al*., 2016; Furukawa *et al*., 2013; Zhao & Miao, 2024; Wang *et al*., 2024; Gomez-Romero *et al*., 2024; Wang & Didier, 2020). The magnetic properties of CPs are of particular interest, and can be combined with other physical characteristics, to produce multifunctional materials. This multifunctionality is enabled by the presence of both organic molecules and metal cations in the same network and creates a plethora of opportunities for the development of novel smart materials (Coronado & Mínguez Espallargas, 2013; Luo *et al*., 2016; Coronado*,* 2020; Verdaguer & Gleizes, 2020; Liu *et al*. 2022). Aperiodic CPs are of growing interest to the crystallographic community as, despite being long-range ordered, they lack the three-dimensional periodicity which underlies many of the fundamental assumptions of diffraction analysis. Modulated crystals are an important class of aperiodic crystals (van Smaalen, 2004). A structure is modulated where the average translational symmetry is disrupted by the introduction of an additional periodic function. The modulation can describe the atomic displacements or occupations, for structural modulations (Pinheiro & Abakumov, 2015; Janssen & Janner, 2014), where the periodicity of the modulations exceeds that of the average structure, i.e. the recurrent part of the structure is larger than the unit cell of the parent structure. If the modulation periodicity can be described by a rational fraction, the structure is commensurately modulated. If an irrational value is necessary, the compound is incommensurately modulated (van Smaalen, 2004).

The signature of modulated phases are satellite reflections in their diffraction patterns: reflections that cannot be indexed by a three-dimensional space group and separate from the main Bragg reflection by a defined spacing. From the satellite reflections, the modulation periodicity, described by the wavevector **q**, can be calculated. The driving force behind the modulated phases lies in unresolved frustration (Dzyabchenko & Scheraga, 2004; Herbstein, 2005; Schönleber, 2011). Explored examples of mechanisms which have induced modulation include cooperative Jahn-Teller distortions (Noda *et al*., 1978), inter/intra-molecular steric constraints (Bakus *et al*., 2013) and hydrogen bonding (Cañadillas-Delgado *et al*., 2019). In each case, the competing interactions result in the loss of translational symmetry between average unit cells.

Reports of molecular frameworks that exhibit modulated phases are still limited (Aroyo *et al*., 2011; Aroyo *et al*., 2006). This is particularly remarkable since weak interactions commonly observed in CPs, including hydrogen bonding, dipole-dipole interactions, and $\pi$-stacking, are equivalent to the forces that usually generate aperiodic systems (Pinheiro & Abakumov, 2015). This would suggest that many of the published compounds might have unreported modulated phases (Oppenheim *et al*, 2020). However, the study of these systems is of potential interest because the intrinsic properties of the material, such as phonon, electric, magnetic, photonic or molecular transport properties, are likely to be different from those of periodic materials (Janssen & Janner, 2014; Poddubny & Ivchenko, 2010; de Regt *et al*, 1995; Vardeny *et al*, 2013). This is important in the sense that in many cases a more comprehensive review of the literature is necessary to identify novel structure-property relationships (Allendorf *et al*, 2021).

The methylammonium metal formates, $[CH_3NH_3]M(HCOO)_3$, M = Co and Ni (Cañadillas-Delgado *et al*., 2019; Cañadillas-Delgado *et al*., 2020) are more unusual, as both undergo phase transitions from unmodulated to incommensurately modulated structures on cooling. Although isomorphous at ambient temperature, the series of temperature-induced phase transitions exhibited by the two compounds are not equivalent (Figure 1).

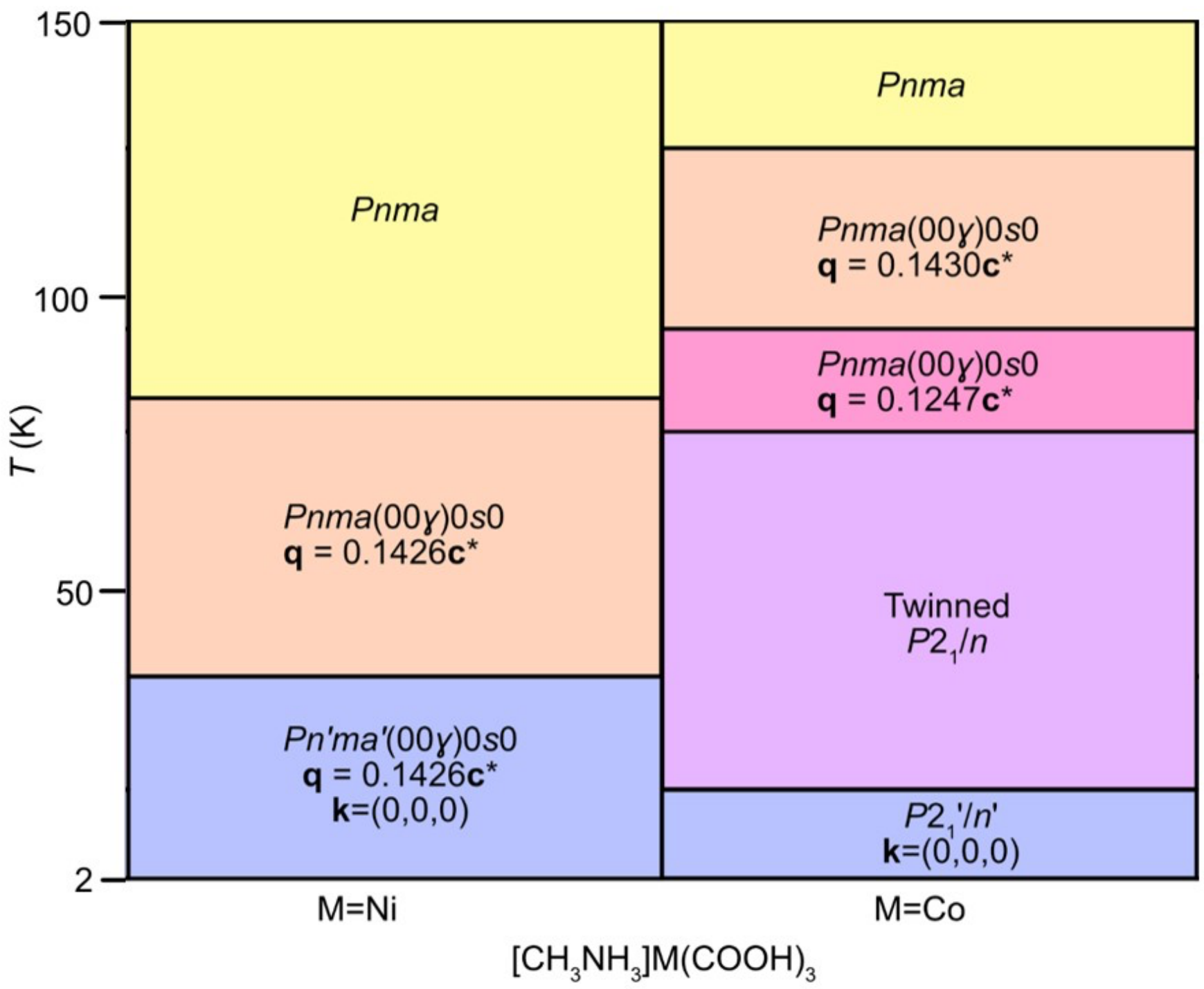

**Figure 1** The temperature dependent phase evolution for a single crystal of $[CH_3NH_3]Co(HCOO)_3$ and $[CH_3NH_3]Ni(HCOO)_3$ from 2 to 150 K. Above 150 K to ambient temperature, there are no further phase transitions observed. The lowest temperature transitions for each metal (blue) correspond to magnetic ordering.

At ambient temperature $[CH_3NH_3]M(HCOO)_3$, M = Co and Ni, crystallise in the orthorhombic space group *Pnma* (Boča *et al*., 2004), adopting a perovskite like structure $ABX_3$, with the metal atoms occupying the B-site, the methylammonium cation located in the A-site, and the formate ligand serving as a linker between the metal atoms at the X-site. Each metal site (Wyckoff site 4*b*) is octahedrally coordinated and is bridged by formate anions in the *anti-anti* coordination mode to form a three-dimensional framework. The carbon and nitrogen atoms of the methylammonium cations are positioned at (x, 0.25, z) and (x, 0.75, z), within the voids of the framework. On cooling, $[CH_3NH_3]Co(HCOO)_3$ experiences its first phase transition at approximately 128 K to the superspace group *Pnma*(00*γ*)0*s*0 (Cañadillas-Delgado *et al*., 2019). In this phase, there is a modulated unit cell, with the wavevector **q** = 0.1430(2)**c***, describing a modulation length of 6.99 times that of the average *c* axis. At 96 K, a second transition is observed where there is a change in the wavevector **q** = 0.1247(2)**c***, however the symmetry of the crystal remains the same, *Pnma*(00*γ*)0*s*0. In this phase, the modulation length is approximately 7.92 times larger than the average unit cell. For both the modulated phases, the atom displacement occurs predominately along the *b* axis, with the amplitude displacements larger for the second modulated phase. Below 78 K a fourth phase is obtained, a twinned, non-modulated monoclinic structure with $P2_1/n$ symmetry (Mazzuca *et al*., 2018), with two domains, contributing 50 % to the total intensity, related by a rotation of 180° around the orthorhombic *c** axis. The onset of long-range magnetic ordering with weak ferromagnetic behaviour is observed at 16 K, with the ferromagnetic component along the *c* axis (Gómez-Aguirre *et al*., 2016; Pato-Doldán *et al*. 2016; Ding *et al.,* 2023). In the single crystal study, only the monoclinic $P2_1'/n'$ magnetic structure is observed (Cañadillas-Delgado *et al*., 2019) (Figure S1a), however, with powder neutron diffraction data a combination of $P2_1'/n'$ and *Pn'ma'* magnetic phases are reported (Mazzuca *et al*., 2018).

In comparison, the nickel analogue remains in the non-modulated *Pnma* space group on cooling until 84 K (Cañadillas-Delgado *et al*., 2020). Below this temperature, it adopts the superspace group *Pnma*(00*γ*)0*s*0 with **q** = 0.1426(2)**c***. $[CH_3NH_3]Ni(HCOO)_3$ remains in this incommensurately modulated phase until the onset of long-range magnetic ordering at 34 K (Pato-Doldán *et al*. 2016), where the compound orders in the magnetic superspace group *Pn'ma'*(00*γ*)0*s*0. Here, the phase is described as a proper incommensurate magnetic structure, as the magnetic moments network presents a modulation due to the occurrence of incommensurate magnetic modes. Consequently, in this phase there is the coexistence of an incommensurately modulated nuclear and magnetic structure. The moments are orientated primarily along the *c* axis, with an uncompensated contribution in the *b* direction (Figure S1b and c). The modulation of the moments occurs as static librations in the *ac* plane. The stimulus driving the modulated phase transitions for both the formate compounds is the hydrogen bond network between the $NH_3$ hydrogen atoms of the methylammonium cations and the oxygen atoms of the formate ligands. In the ambient temperature phase, two of the hydrogen atoms participate in hydrogen bond interactions, whilst the third is too far in proximity to interact with the neighbouring oxygen atoms (Figure S2a). In the low temperature $P2_1/n$ phase, adopted by $[CH_3NH_3]Co(HCOO)_3$, a third permanent hydrogen bond is present (Figure S2b). The methylammonium cation, which is positioned along a mirror plane in the orthorhombic phase, rotates in the monoclinic phase breaking the symmetry element. H1 (atom labels as in Figure S2d), which was originally equidistance to O3 and O3*a* (with $a = x, -y + 3/2, z$), is now closer to one of the oxygen atoms, establishing a hydrogen bond interaction. In the modulated phases, H1 is close enough in distance to interact with the oxygen atoms of the formate ligand. However, there are two competing hydrogen bond acceptors (O3 and O3*a*, Figure S2c), equidistance from H1 in the average structure. This frustration distorts the structure and results in the modulated displacement of all the atoms. Consequently, at a given point in the structure O3 will be closer in distance to H1, yet, in other areas along the modulation the N1−H1···O3*a* distance will be shorter.

In a recent work on the controlling of the modulated phases (Li *et al*., 2020), the authors analyse a metal organic framework material, MFM-520, which displays a reversible periodic-to-aperiodic structural transition through host-guest interaction. The dehydrated phase presents an aperiodic structure with translational symmetry in (3+2)D space, which changes to a periodic phase when $H_2O$ molecules are incorporated in the pores of the structure. Subsequent substitution of $H_2O$ molecules with $CO_2$ and $SO_2$ revealed that, while $CO_2$ exerts minimal structural influence, $SO_2$ can also induce modulation in the structure. This study motivated us to investigate the feasibility of combining Ni and Co in the perovskite B-site to explore the sensitivity of the modulated phase transitions and magnetic characteristics of the formate compounds.

In the present work, solid solutions have been synthesised from the methylammonium metal formates, $[CH_3NH_3]Co_xNi_{1-x}(HCOO)_3$, with $x = 0.25$ (**1**), 0.50 (**2**) and 0.75 (**3**), and their structures and magnetic properties have been analysed through single-crystal neutron diffraction and magnetometry studies. All three compounds were studied through Laue experiments on CYCLOPS and monochromatic measurements on D9 instruments, and a deep study of the modulated phases and magnetic structure was done on compound **2** on D19 monochromatic diffractometer. Measurements on a Quantum Design Magnetic Property Measurements System (MPMS) and a Superconducting Quantum Interference Device (SQUID) magnetometer were carried out for all compounds.

## 2. Results

### 2.1. Synthesis and ambient structure

Methylammonium formate solid solutions were synthesized under solvothermal conditions from aqueous solutions of $NiCl_2·6H_2O$, $CoCl_2·6H_2O$, $CH_3NH_3Cl$ and NaHCOO in stoichiometric quantities following the method previously reported (Mazzuca *et al*., 2018; Wang *et al*., 2004). The solutions were heated at 413 K (140 °C) for 72 h before cooling, within the sealed autoclave, to room temperature. This yielded $mm^3$ sized crystals of **1** (dark green), **2** (dark green) and **3** (dark red).

Single crystal X-ray diffraction measurement reveal that the compounds are isostructural to the end-member methylammonium formate compounds $[CH_3NH_3]Co(HCOO)_3$ and $[CH_3NH_3]Ni(HCOO)_3$ (Boča *et al*., 2004; Pato-Doldán *et al*. 2016) (Figure 2). Neutron diffraction studies were performed using the D9 diffractometer at the ILL to obtain more precise information on the metal site orderings and the cobalt and nickel ratio in each crystal (Geers & Cañadillas-Delgado, 2021b; Cañadillas-Delgado *et al*., 2023). Room temperature diffraction data were collected for single crystals of **1** (2×1.5×1.5 $mm^3$), **2** (3×2×2 $mm^3$) and **3** (1.5×1.5×1 $mm^3$). The cobalt and nickel occupations were refined and were constrained such that the overall site occupancy was 1 and while the site positions and anisotropic displacement parameters for both species were constrained to be equal.

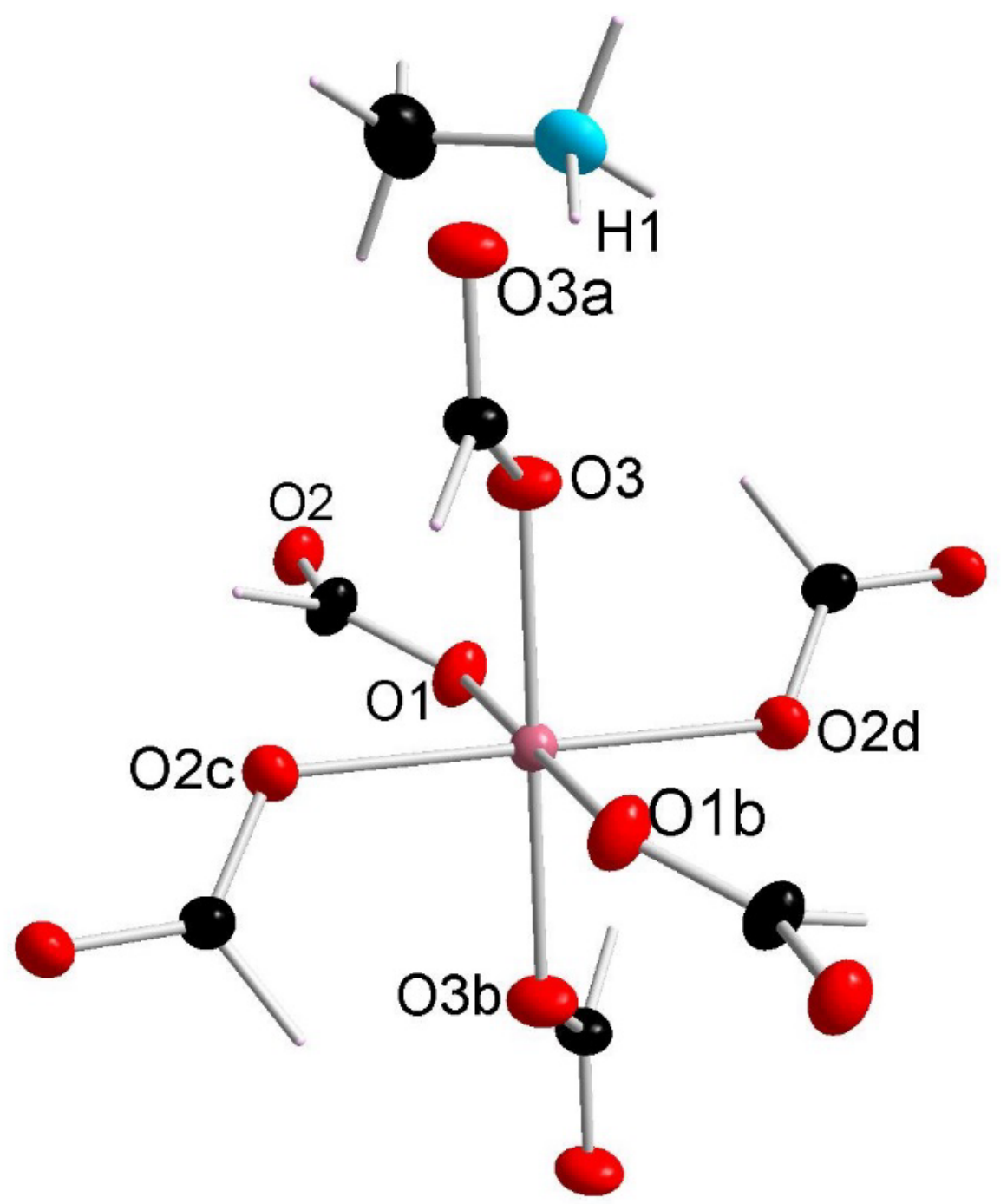

**Figure 2** The connectivity of the metal-formate framework for compound **2** measured with the D9 diffractometer (ILL) at ambient temperature. Compounds **1** and **3** are isostructural. The ellipsoids are drawn with 50% probability and the hydrogen atoms are represented as sticks for clarity. *Colour code*: Metal = purple, O = red, C = black, N = blue. *Symmetry code*: *a* = *x*, -*y*+3/2, *z*; *b* = -*x*+1, -*y*+1, -*z*+2; *c* = -*x*+1/2, -*y*+1, *z*+1/2; *d* = *x*+1/2, *y*, -*z*+3/2

The neutron diffraction data shows that there is no ordering of the metals. The symmetry for the solid solutions are the same as for the end members, *Pnma*, with no superlattice reflections observed or reflections corresponding to systematic absences in the *Pnma* space group. The metal content for each crystal corresponds to $x = 0.297(9)$, 0.526(8) and 0.765(20) for **1**, **2** and **3**, respectively. These values are close to the stoichiometric quantities of metal chlorides used in the synthesis.

### 2.2. Magnetometry

Field-cooled (FC) and zero-field-cooled (ZFC) susceptibility measurements were carried out on microcrystalline samples using an applied field of 100 Oe over the temperature range 2 − 300 K. The extent of field induced magnetisation was also explored at 2 K between −5.00(1) and 5.00(1) T.

The susceptibility data for compound **1** ($Co_{25}Ni_{75}$) indicate an ordering temperature $T_C = 28.5(5)$ K (Figure 3a). From the plot of $\chi T$, the effective moment is calculated to be 3.936(2) $\mu_B$ in comparison to $\mu_{spin\ only} = 3.09\ \mu_B$ and the Curie-Weiss temperature $\theta_{CW} = -70.8(7)$ K ($150 < T < 300$ K). The high-temperature Curie constant, $C$, extracted from the Curie-Weiss fit in the range from 150 to 300 K is = 2.432(7) emu K mol$^{-1}$ (Figure S3a) which is larger than the spin only value $C_{spin\ only} = 1.21$ emu K mol$^{-1}$. This discrepancy in the Curie constant with respect to the spin-only value indicates strong orbital contributions, as is also reflected in the pure Co compound (Pato-Doldán *et al*. 2016), and is also present in the other solid solutions (**2** and **3**). Both $Ni^{2+}$ and $Co^{2+}$ ions have large orbital contributions to the magnetic moment, meaning that in all these formate compounds $\theta_{CW}$ varied greatly with the temperature range used to perform the fit. The isothermal data, reveal hysteresis at all magnetic fields measured for **1** (Figure S4a). It exhibits a remnant magnetisation $M_{rem} = 0.015(1)\ \mu_B$ per metal and a coercive field of $H_C = 0.15(1)$ T. The magnetisation does not reach saturation, $M_{sat}$ = 1.125(1) $\mu_B$ per metal, with $M_{5\,T} = 0.158(1)\ \mu_B$ per metal at 5.00(1) T ($M_{5\,T}/M_{sat.} = 0.140$).

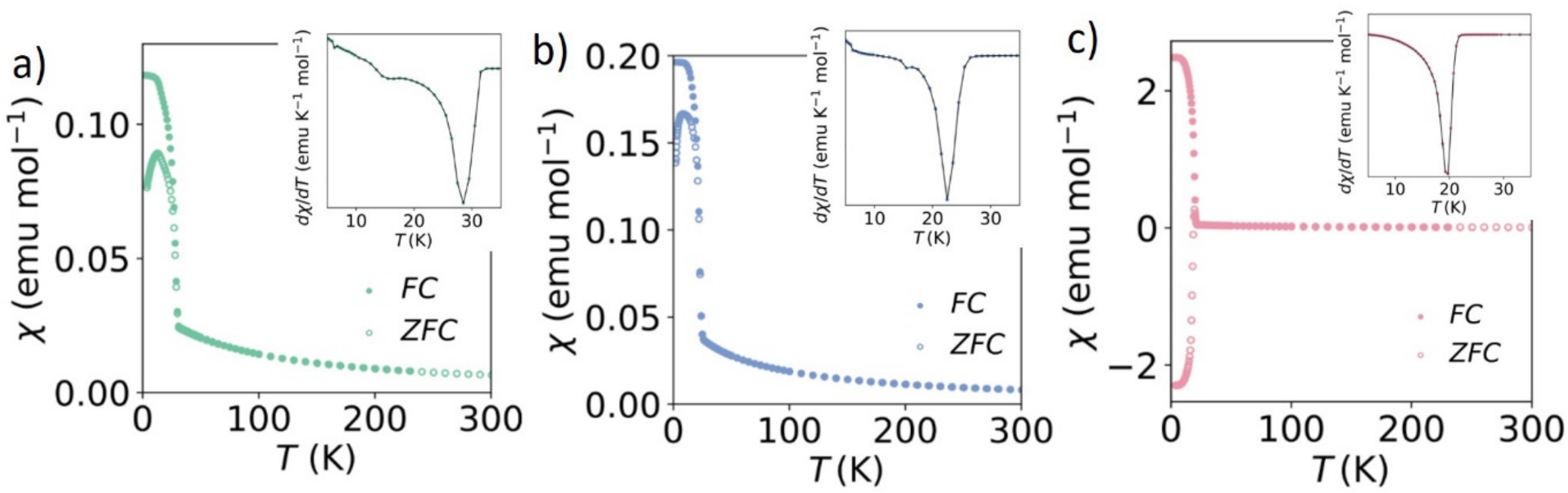

**Figure 3** Field-cooled (FC) and zero field-cooled (ZFC) susceptibilities for compounds **1** (a), **2** (b) and **3** (c). The insets correspond with the derivative of the susceptibility.

For compound **2** ($Co_{50}Ni_{50}$), the field-cooled and zero-field-cooled susceptibilities diverge at the ordering temperature, $T_C$ =22.5(7) K (Figure 3b). The effective magnetic moment $\mu_{eff.} = 4.436(2)\ \mu_B$ compared to $\mu_{spin\ only} = 3.35\ \mu_B$. Fitting the inverse susceptibility to the Curie-Weiss Law we obtain $\theta_{CW} = -56.3(1)$ K and $C = 2.938(1)$ emu K mol$^{-1}$ in the range $150 < T < 300$ K, which is significantly higher than the spin only value $C_{spin\ only} = 1.44$ emu K mol$^{-1}$ (Figure S3b). Hysteresis is observed in the magnetisation data of **2**, which closes at 5.0(2) T (Figure S4b). The remnant magnetisation is $M_{rem}$

= 0.027(1) $\mu_B$ per metal, with a coercive field of $H_C$ = 0.30(1) T. The largest magnetisation measured is $M_{5\,T}$ = 0.218(1) $\mu_B$ per metal, which is lower than the saturation value $M_{sat}$ = 1.25 $\mu_B$ per metal ($M_{5\,T}/M_{sat}$ = 0.175).

As compound **3** ($Co_{75}Ni_{25}$) is cooled, the field-cooled and zero-field-cooled susceptibilities split at the ordering temperature, $T_C$ = 19.7(5) K (Figure 3c). Similarly, the effective moment, $\mu_{eff}$ = 4.578(2) $\mu_B$, is larger than the spin only value 3.61 $\mu_B$. The Curie-Weiss temperature, obtained from a high temperature fit of the data 150 < $T$ < 300 K, is $\theta_{CW}$ = −43.6(5) K, and $C$ = 2.997(3) emu K mol$^{-1}$ in the same temperature range, which is greater than the spin only value for the average magnetic site $C_{spin\ only}$ = 1.66 emu K mol$^{-1}$ (Figure S3c). The isothermal magnetisation measurements for compound **3** show that as the field is increased, hysteresis is observed up to 3.04(1) T, where the magnetisation $M$ = 0.219(1) $\mu_B$ per metal (Figure S4c). The remnant magnetisation is $M_{rem.}$ = 0.045(3) $\mu_B$ per metal, and the hysteresis has a coercive field of $H_C$ = 0.45(1) T. At the largest field measured, 5.00(1) T, the magnetisation reaches $M_{5\,T}$ = 0.337(2) $\mu_B$ per metal, with no signs of a plateau. The degree of saturation is $M_{5\,T}/M_{sat}$ = 0.245, where the saturation value $M_{sat}$ = 1.375 $\mu_B$ per metal.

For all three compounds, the combination of a negative $\theta_{CW}$ and hysteresis in the isothermal data suggest weak ferromagnetic behaviour, in agreement with the magnetic ordering of the parent compounds. If the crystals were biphasic, rather than true solid solutions, it would be expected that there are two ordering temperatures observed in the magnetometry data, one for the pure Co compound at 16 K and one for the pure Ni compound at 34 K. For the samples measured, essentially one ordering temperature is observed which incrementally increases between the Co and Ni ordering temperatures, dependent on the metal ratios used in the synthesis. This would support the X-ray and neutron diffraction data that there is no metal site ordering and the distribution of Co and Ni is random within the samples.

### 2.3. Temperature dependent structural evolutions

Single crystal Laue neutron diffraction was carried out for compounds **1**, **2** and **3** on CYCLOPS diffractometer at ILL, with a wavelength range of 0.8−3.0 Å on the same crystals used on D9 diffractometer (Geers & Cañadillas-Delgado, 2021a). The samples were heated between 10 and 120 K and the diffractograms were collected with a 3 K range per image. This technique was not used to determine the structures, but to determine the temperatures at which structural phase transitions appear and estimate the **q** vectors of modulated phases.

For the nickel rich compound, **1**, the crystal remains in the *Pnma* space group on cooling until 85(3) K. At this temperature, satellite reflections are visible (Figure S5). The reflections match a calculated diffraction pattern for the *Pnma* space group with **q** = 0.140(5)**c***. These reflections remain as the sample is cooled to 10 K, without observable alteration to the **q** vector. Below 28(3) K, although no new reflections appear in the pattern of the selected orientation, certain reflections increase in intensity (Figure S5e). This observation aligns with the long-range magnetic ordering temperature identified through magnetometry measurements ($T_C$ = 28.5(5) K), indicating a **k** = (0,0,0) propagation vector.

When cooling, compound **2** remains in a non-modulated *Pnma* phase until 96(3) K, at which temperature satellite reflections appear and the main reflection reduces in intensity (Figure S6b). The reflections can be modelled by the *Pnma* space group, with **q** = 0.140(5)**c***. A smooth phase transition can be observed at 59(3) K with a change in distance between the main and satellite reflections as well as the appearance of additional satellite reflections, which increase in intensity down to 33(3) K (Figure S6c). The reflections can be matched to the *Pnma* space group with **q** = 0.120(5)**c***. Below this temperature the position of the main and satellite reflections do not change further. Additional intensity to some reflections can be observed below 25(3) K, in line with the ordering temperature observed from the magnetisation data ($T_C$ = 22.5(7) K, Figure S6d). This implies that the magnetic structure has a propagation vector of **k** = (0,0,0).

Cooling from 120 K, the diffraction pattern for **3** remains unchanged in the non-modulated *Pnma* space group until 98(3) K where the emergence of satellite peaks can be observed (Figure S7b). Below 98(3) K, the main Bragg reflections match a calculated diffraction pattern for the *Pnma* space group and the satellite peaks agree with a **q** vector of about 0.135(5)**c***, which subtly decreases to **q** = 0.125(5)**c*** by 58(3) K. From 58(3) K a slow transition occurs over 14(3) K temperature range (Figure S7c). During the transition, the reflections reduce in intensity and are distributed over a larger pixel area. By 44(3) K, the reflections are again sharp intensities. The reflections no longer have satellite peaks, suggesting this is a non-modulated structure, and twinning can be observed by the appearance of a second Bragg reflection close in proximity to the main reflection (Figure S7d). The reflections fit the monoclinic space group $P2_1/n$, like the low temperature phase of the pure cobalt formate compound. At 18(3) K, certain reflections increase in intensity, indicating the onset of long-range magnetic ordering, which is in agreement with the ordering temperature determined by the magnetometry measurements ($T_C$ = 19.7(5) K). Since the magnetic reflections only appear as additional intensity to nuclear reflections, it likely has a propagation vector of **k** = (0,0,0).

### 2.4. Monochromatic neutron diffraction

The Laue diffraction data show that the cobalt-rich and nickel-rich solid solutions follow structural phase transitions similar to that of the end-member methylammonium formates. However, the evolution of the $Co_{50}Ni_{50}$ (compound **2**) structure did not strictly follow the trend for either the Co or Ni analogue, encouraging a further neutron diffraction experiment to explain the observed behaviour.

A single crystal neutron diffraction experiment was carried out using the D19 diffractometer (ILL), using the same crystal that was used on CYCLOPS and D9 diffractometers (Geers *et al.*, 2021). Data were collected at 2 K and at 10 K intervals between 30 and 100 K. The orientation matrix was obtained for each data set to identify the temperature regimes for each phase. As a result, longer data acquisitions were made at 30 and 70 K. Refinements were carried out at 2, 30 and 70 K to determine the nuclear and magnetic structures of each phase. For the following refinements, the previously calculated metal ratio for this crystal was used (0.526:0.474, Co:Ni) and were fixed during the refinements. A summary of the experimental and crystallographic data can be consulted on Table 1.

**Table 1** Experimental and crystallographic data of compounds **1-3**, measured on D9 and D19 neutron diffractometers refined with JANA2020 software (Petříček *et al.*, 2023). All H-atom parameters were refined for all compounds.

| Chemical formula | $C_4H_9Co_xNi_{1-x}NO_6$ | | | | | |
|---|---|---|---|---|---|---|
| Compound | **1** | **2** | | | | **3** |
| Refined $x$ | 0.297(9) | 0.526(8) | | | | 0.765(2) |
| $M_r$ | 225.9 | 225.9 | | | | 226.0 |
| Z | 4 | 4 | | | | 4 |
| Diffractometer | D9 | D9 | D19 | D19 | D19 | D9 |
| Temperature (K) | RT | RT | 70 | 30 | 2 | RT |
| Space group | *Pnma* | *Pnma* | *Pnma*(00*γ*)0*s*0 | *Pnma*(00*γ*)0*s*0 | *Pnma*(00*γ*)0*s*0 | *Pnma* |
| *a*, Å | 8.358(2) | 8.3506(4) | 8.2052(2) | 8.2003(2) | 8.2010(3) | 8.372(2) |
| *b*, Å | 11.637(3) | 11.6556(8) | 11.5759(3) | 11.5737(3) | 11.5747(8) | 11.705(4) |
| *c*, Å | 8.069(2) | 8.0831(4) | 8.1141(2) | 8.1133(2) | 8.1144(3) | 8.095(2) |
| $V$, Å$^3$ | 784.8(3) | 786.74(8) | 770.70(3) | 770.02(3) | 770.25(7) | 793.2(4) |
| Wavevectors | - | - | **q** = 0.1429(2)**c*** | **q** = 0.1249(2)**c*** | **q** = 0.1249(2)**c*** | - |
| $\rho_{calc}$, mg m$^{-3}$ | 1.9118 | 1.9075 | 1.9472 | 1.9489 | 1.9483 | 1.8925 |
| λ, Å | 0.8348 | 0.8359 | 1.45567 | 1.45567 | 1.45567 | 0.8359 |
| μ, mm$^{-1}$ | 0.005 | 0.006 | 0.011 | 0.011 | 0.011 | 0.008 |
| $R_1$, I > 3s(I) (all) | 0.0354 (0.0574) | 0.0361 (0.0611) | 0.0601 (0.0762) | 0.1189 (0.1315) | 0.1217 (0.1343) | 0.0636 (0.1109) |
| $wR_2$, I > 3s(I) (all) | 0.1048 (0.1067) | 0.0531 (0.0645) | 0.2185 (0.2321) | 0.1899 (0.2034) | 0.1802 (0.1901) | 0.1235 (0.1525) |
| No. of parameters | 107 | 107 | 164 | 449 | 452 | 107 |
| Independent reflections | 737 | 1178 | 3291 | 3346 | 3346 | 528 |
| No. of main reflections | - | - | 682 | 694 | 694 | - |
| No. of 1st-order satellite reflections | - | - | 1230 | 1250 | 1250 | - |
| No. of 2nd-order satellite reflections | - | - | 1379 | 1402 | 1402 | - |

At 96(3) K, compound **2** undergoes a phase transition from the non-modulated, orthorhombic space group, *Pnma* to a modulated structure. Integration of the data at 70 K found that the compound adopts the modulated superspace group *Pnma*(00*γ*)0*s*0 with a modulation vector **q** = 0.1429(2)**c***.

Refinements of the amplitude displacements for the atoms at 70 K reveal that the site displacement is largest along the *b* direction. The modulation for each atom was refined independently to find the displacements. For the metal site, the displacement is described by a sine term only (restricted by symmetry, Table 2), with a maximum displacement from its average position of 0.235(3) Å in the *b* direction (Figure S8). Asynchronous modulations of the atoms result in the modulation of the bond lengths, however, these variations are of two orders of magnitude smaller than the displacements experienced for the atoms.

**Table 2** The amplitude displacements for the sine term of the first order of the harmonics in the Fourier series of the metal site for compound **2**.

| | 70 K | 30 K | 2 K |
|---|---|---|---|
| *x* | 0.00216(13) | 0.00283(19) | 0.00235(19) |
| *y* | 0.02033(15) | 0.0322(2) | 0.0335(2) |
| *z* | 0.00009(14) | 0.00020(19) | 0.0001(2) |

The hydrogen bond interactions between the donor N−H methylammonium and acceptor O atoms vary in distance as the structure modulates. The H2· · ·O2 interactions (atoms labelled as in Figure S2d) vary in the range from 1.807(3) Å to 1.830(3) Å, indicating that both atoms preserve an appropriate distance to maintain hydrogen bonding in all regions of the crystal. H1···O3 and H1···O3*a* distances varies between 2.028(3) Å and 2.265(3) Å. This variation of distances gives rise to flip-flop behaviour, so that in some regions of the crystal a possible H-bond is established with oxygen O3, in other regions a minimum distance is established with oxygen O3*a* and in other regions the distances to O3 and O3*a* are sufficiently long to dismiss the hydrogen bond.

Decreasing in temperature, a second phase transition is observed between 40 and 30 K. This is slightly lower than the transition that was observed in the Laue diffraction data at 59(3) K. Potentially this discrepancy in the transition temperature is as a result of the temperature continually increasing as a function of time for the Laue diffraction measurements, whereas during these data collections the sample was allocated time to stabilise at each temperature point before starting the data collection.

At 30 K, compound **2** presents a crystal structure refined in the superspace group *Pnma*(00*γ*)0*s*0, with a modulation vector, **q** = 0.1249(2)**c***. Although new satellite reflections appeared in the Laue measurement below 59(3) K, monochromatic measurements revealed that all the satellite reflections in this phase are first and second order, like in the previous measurement at 70 K. This contrasts with the measurement of the pure Co compound where third order satellites were visible below 90 K, i.e. the temperature corresponding to the change of the wavevector from **q** = 0.1430(2)**c*** to **q** = 0.1247(2)**c***. The lack of third-order satellites is likely due to compound **2** being a solid solution which could tend to reduces homogeneity and the overall crystallinity of the sample. Like in the previous described phase, the metal site positions modulate with the largest contribution along the *b* direction (Table 2). This results in a maximum displacement along the *b* axis from its average position of 0.373(3) Å (Figure S8).

The M−O bond lengths exhibit larger distortions from the average value at 30 K compared with 70 K phase. This is particularly significant for M−O1 and M−O3 (Table S2). M−O3 coordinate to the metal sites along the *b* direction, with a maximum deviation from the average bond length of 0.069(5) Å. The M−O1 bonds are located in the *ab* plane with a maximum bond length variation of 0.060(3) Å. The difference in maximum bond length for M−O2 compared to its average value is an order of magnitude smaller, 0.007(3) Å. The modulated distance H2···O2 ranges from 1.784(4) Å to 1.863(3) Å, indicating that the interaction is present throughout the crystal. The H1···O3 and H1···O3*a* modulations alternate, resulting in minimum and maximum distances at different points in the structure, as in the previous phase (Figure S9b).

On further cooling to 2 K, additional reflections were observed, indicating the onset of magnetic ordering. The presence of the satellite reflections implies that the structure remains in a modulated phase. However, initially it was unclear if the modulation arises from only the nuclear structure (improper incommensurate magnetic structure) or from a combination of the nuclear and magnetic structure (proper incommensurate magnetic structure). Refinements were carried out both where the Fourier components for the magnetic moments were refined and where the modulations were fixed to zero for the moments. As equivalent refinement statistics were obtained for both models, and there was no evidence of additional intensities in the satellite reflections, but instead only an increase in intensities of the main reflections, it was concluded the structure adopts an improper modulated magnetic structure.

From indexing the magnetic Bragg reflections, the propagation vector was determined to be **k** = (0,0,0). **2** orders with the magnetic superspace group *Pn'ma'*(00$\gamma$)0*s*0 with **q** = 0.1249(2)**c***. This model allows for a weak ferromagnetic arrangement of the moment, in agreement with the susceptibility data. The moment was fixed to have a magnitude of 2.50(2) $\mu_B$, an average of high-spin $Co^{2+}$ and $Ni^{2+}$ moments. Each nearest neighbour, through M−OCO−M superexchange pathways, has weak ferromagnetic correlations, caused by a canting of the moments along the *b* axis at an angle of ca. 104° (0.6(2) $\mu_B$) with respect to this axis, (Figure 4). Since the magnetic component of the structure is non-modulated, the orientation and size of the moment does not change throughout the structure. The nuclear component, however, remains modulated. The metal site modulates and, accordingly, the position of the moment is displaced from the average structure, however, the magnitude and direction of the moments are not varied. The greatest displacement of the atom sites is along the *b* axis (Table 2), with a maximum displacement of 0.388(6) Å (Figure S8, blue line).

The length of the *c*-axis, as well as the unit cell volume in the modulated phases, is 7 and 8 times larger than in the commensurate orthorhombic phase (*Pnma* phase). In these cases, a refinement in a non-modulated phase is possible, however the number of parameters becomes too large, so that the number of measured reflections is not sufficient and the refinement becomes unstable. This has already been tested in the pure Co compound (Cañadillas-Delgado *et al.*, 2019), where it is observed that the space group would become $P2_12_12_1$, with a number of atoms in the asymmetric unit equal to 147, compared to the 14 atoms of the asymmetric unit in the supergroup.

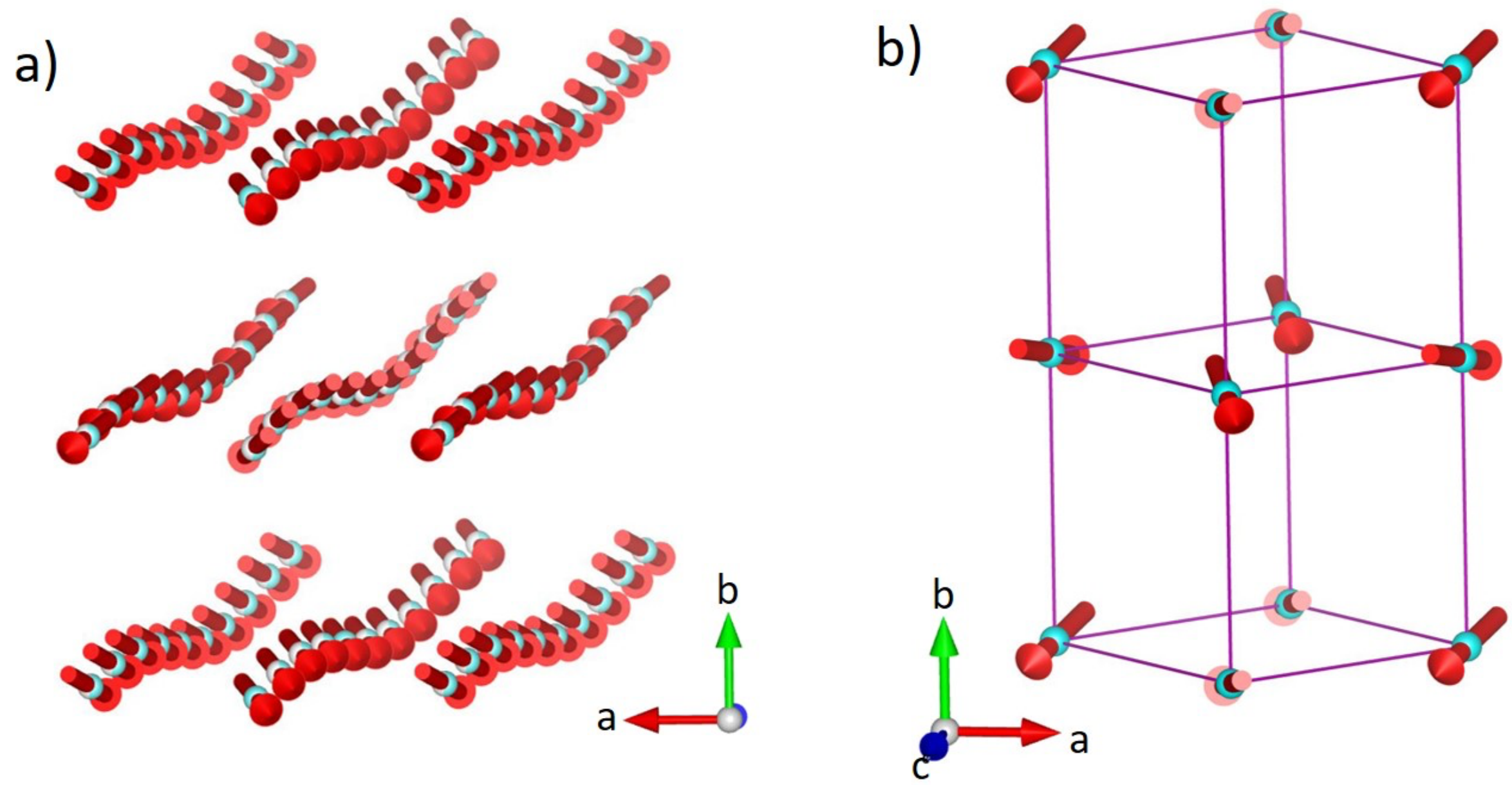

**Figure 4** The magnetic structure of compound **2** at 2 K measured with the D19 diffractometer (ILL). a) The position of the metal sites and the magnetic moments are represented in a supercell which is 10 times larger than the average unit cell to include at least a full modulation period. b) The average magnetic unit cell, showing the ordering of the moments without the modulation of the atom sites. The $[M(HCOO)_3]^-$ framework is displayed as a wireframe (purple lines) and the methylammonium cations and hydrogen atoms have been removed for clarity. The moments are tilted, with an uncompensated moment along the *b* axis.

### 3. Discussion

The three solid solutions of $[CH_3NH_3]Co_xNi_{1-x}(HCOO)_3$, $x$ =0.25 (**1**), 0.50 (**2**) and 0.75 (**3**), show intermediate behaviours of their structural evolutions and magnetic properties compared to their two end members $[CH_3NH_3]Co(HCOO)_3$ and $[CH_3NH_3]Ni(HCOO)_3$. The nickel-rich compound **1** undergoes one non-modulated to modulated phase transition, which remains until 2 K. This behaviour follows the transitions observed by the Ni analogue. On the other hand, compound **3** exhibits structural phase transitions similar to that of the Co analogue, transitioning through modulated phases before adopting a twinned non-modulated structure by 44(3) K (Figure 5). The wavevector of the modulated phase just below 98(3) K, **q** = 0.135(5)**c***, corresponds to an incommensurate modulated structure, rather than near commensurate, as in the rest of compounds. It presents smooth phase transitions in wide ranges of temperatures, together with a distribution of the reflections over a large pixel area in the Laue measurements, that would imply the presence of several domains in the sample. The coexistence of phases appears recurrently in all compounds of this family suggesting that the small energy barrier between phases could be easily overcome by external stimuli such as external pressure. Recently it has been studied the effect of external pressure on the parent compound $[CH_3NH_3]Co(HCOO)_3$, by using high pressure powder X-ray diffraction and Raman spectroscopy (Zhou *et al*, 2023). The increase in pressure at room temperature, gives rise to a phase transition from the orthorhombic *Pnma* to a monoclinic phase, at approximately 6.13 GPa. This study indicates that high pressure can profoundly alter the crystal structure and magnetic properties of these compounds, implying that this external stimulus can also serve to control also the phase transition from the modulate structure at low temperature.

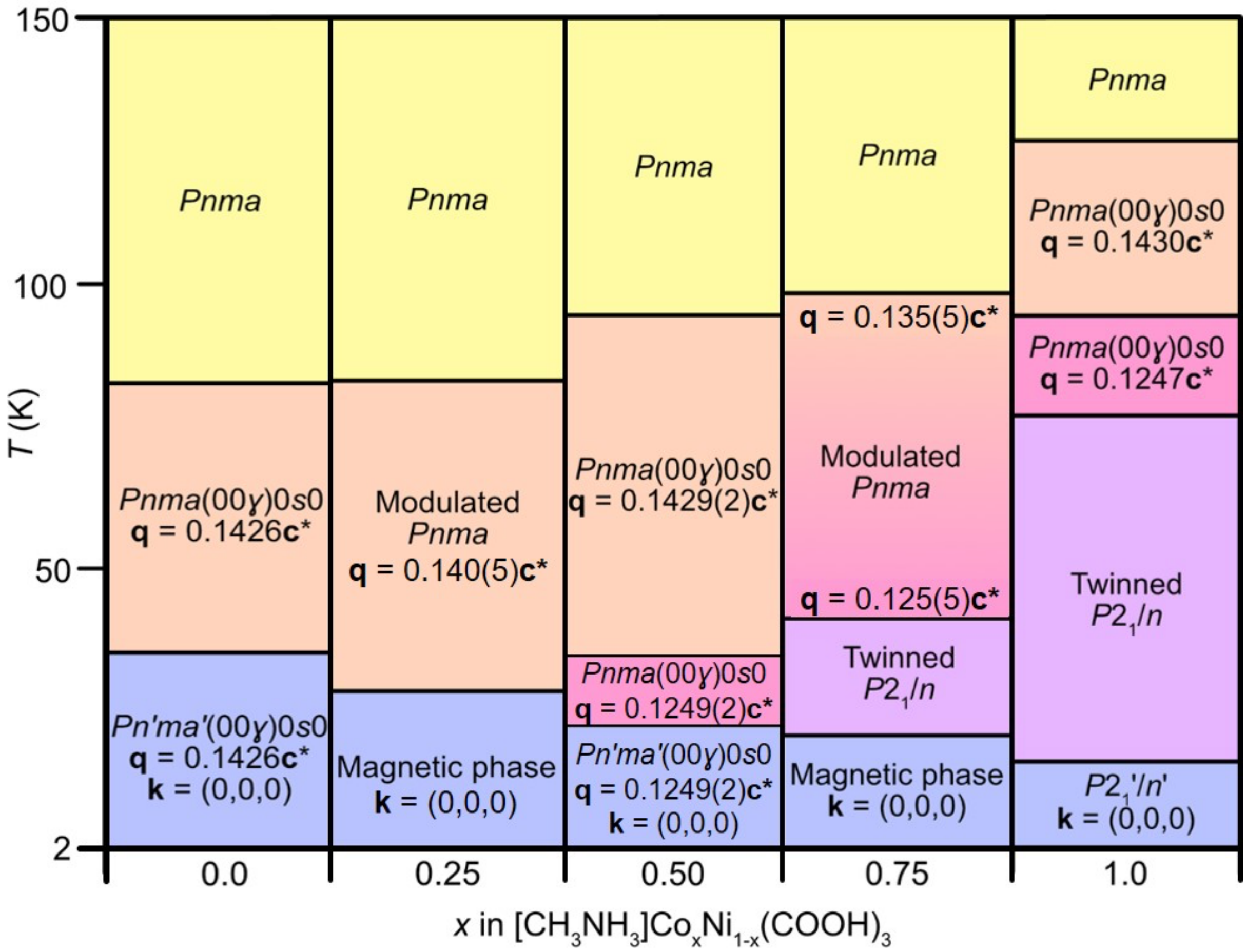

**Figure 5** Summary of the structural and magnetic temperature dependent phase transitions in single crystal samples of $[CH_3NH_3]Co_xNi_{1-x}(HCOO)_3$ with $x$ = 0, 0.25 (**1**), 0.5 (**2**), 0.75 (**3**) and 1. For compounds **1** and **3** transition temperatures were obtained from neutron Laue diffraction measurements, and from monochromatic neutron diffraction measurements for compound **2**. The temperature of the magnetic order (phases represented in blue) are obtained from the magnetometry data.

The structural behaviour of **2** exhibits similarities to both the Co and Ni parent compounds, yet the series of phase transitions do not follow either compound directly. The first two phase transitions, between the non-modulated phase and a modulated phase with **q** = 0.1429(2)**c***, followed by an isomorphous phase transition to a structure with **q** = 0.1249(2)**c***, resembles that of the Co analogue, although occurring at lower temperatures. In comparison to the Co analogue, which undergoes a transition to a twinned non-modulated monoclinic structure, that is retained with the onset of magnetic ordering, compound **2** does not exhibit a low-temperature non-modulated phase. It magnetically orders in the superspace group *Pn'ma'*(00*γ*)0*s*0 with **q** = 0.1249(2)**c*** and **k** = (0,0,0). The magnetic symmetry is similar to that of the Ni analogue, although with a smaller modulation vector and only the nuclear structure that contributes to the modulations, the magnetic ordering is non-modulated.

It is noteworthy that Laue measurements reveal a significant increase in the temperature range at which modulated structures manifest in solid solution compounds, in contrast to pure nickel and cobalt compounds. Specifically, while the temperature range spans 82 K and 50 K for pure nickel and cobalt compounds respectively, compounds **1**, **2**, and **3** exhibit an extended temperature range reaching approximately 83 K, 94 K, and 54 K respectively. This suggests that doping the samples increases frustration in the structure, leading to the stabilization of modulated structures.

From the low temperature monochromatic neutron diffraction data, it can be extracted that the mechanism inducing the modulated phase transitions in **2** is the competition of the hydrogen bonding interactions, akin to its parent compounds. Although hydrogen bonding might not be essential in halide perovskites, research demonstrates that it is approximately three times more robust in formate perovskites (Svane *et al*., 2017). The H2· · ·O2 distance remains at values close to its average value, with only small deviations of up to ±0.040(4) Å. H1· · ·O3, which denotes the hydrogen bonding between the formate oxygens along the *b* axis, shows alternating distances between H1· · ·O3 and H1· · ·O3*a*. There are certain zones in the structure where the H1· · ·O3 atoms have a shorter separation, whereas at other points, H1· · ·O3*a* has the shorter contact. This trend is observed at all three temperatures, driving the modulated phases. Comparing with the parent compounds, there is no clear trend between the changes in the hydrogen bond distances and either the modulated structure that is adopted, or the composition of the compound (Figure 6a).

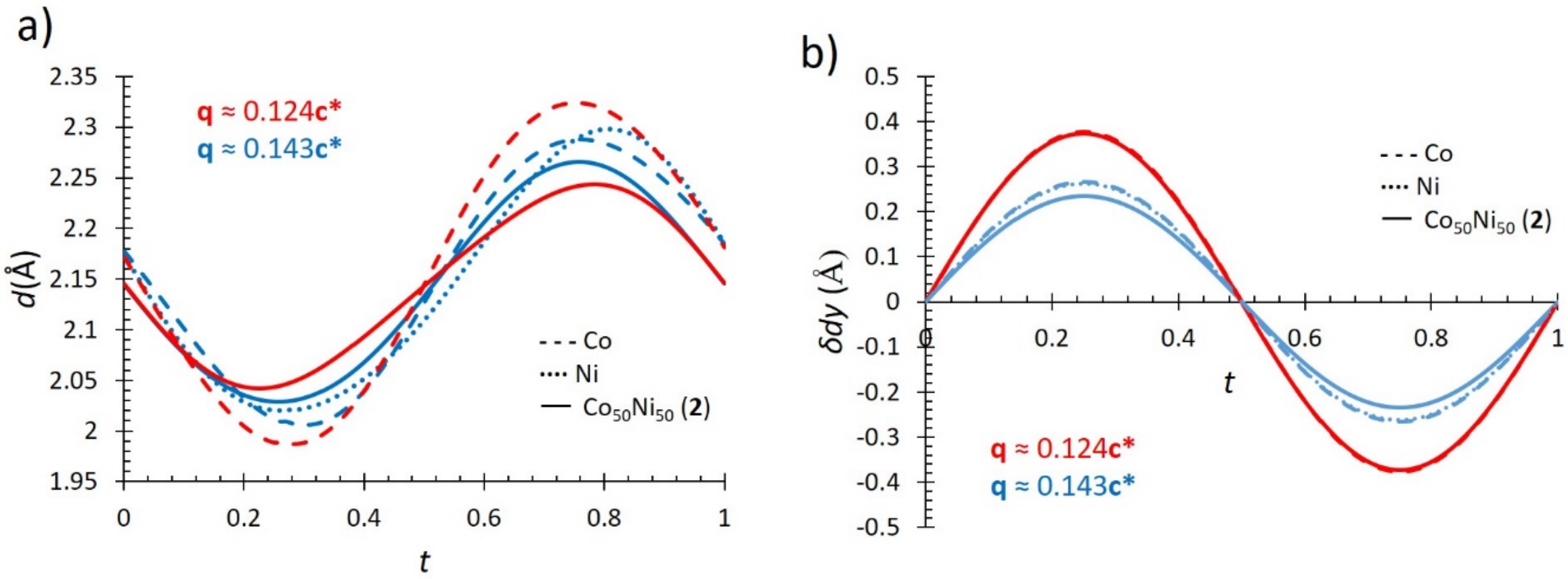

**Figure 6** Summary of the a) H1···O3 bond distances and b) metal displacement along *b* axis for $[CH_3NH_3]Co(HCOO)_3$ at 106 K (blue) and 86 K (red) (dash lines) (Cañadillas-Delgado *et al*, 2019), $[CH_3NH_3]Ni(HCOO)_3$ at 40 K (blue dotted lines) (Cañadillas-Delgado *et al*, 2020) and compound **2** at 70 K (blue) and 30 K (red) (solid lines). Red lines highlight the structures with the approximate wavevector **q** = 0.124**c***  and the blue lines for the approximate wavevector **q** = 0.143**c***.

The modulations are triggered by the hydrogen bond competition, however, the changes in the modulation vector can be observed by the resultant magnitude of displacement of the atom sites from the average structure. This can be followed through the *y* amplitude displacements of the metal site (Figure 6b). There is a division between displacements observed for the shorter modulation length (**q** ≈ 0.143**c***, blue lines) and the atom displacements observed for the larger modulation length (**q** ≈ 0.124**c***, red lines). This is more quantitatively conveyed in amplitude displacements for the sine term of the first order of the harmonics in the Fourier series along *y* for the metal atoms (Table 3). At 70 K, at which temperature **2** has the modulation vector **q** = 0.1429(2)**c***, *y* = 0.02033(15), which is similar to the displacement observed for the Co, *y* = 0.0229 (106 K), and Ni analogues, *y* = 0.02274 (40 K) (Cañadillas-Delgado *et al*., 2019; Cañadillas-Delgado *et al*., 2020). The phase transition to the longer modulation length (**q** = 0.1249(2)**c***) coincides with increasing amplitude displacements in the metal site: for **2** *y* = 0.0322(2) (30 K), compared to the Co analogue *y* = 0.0322(5) (86 K). It is proposed that the Co analogue undergoes its final structural transition to a monoclinic non-modulated phase as the continual increases in the amplitude displacements with temperature eventually result in a division into two non-modulated domains (Cañadillas-Delgado *et al*., 2019). It is possible that the shorter Ni−O bond lengths limit the atom displacement, preventing larger atomic displacement values from being reached. The M−O bond distances at 30 K for **2** are intermediate of the Co−O and Ni−O values, as expected for a solid solution (Table 4) (Lee *et al*., 2016; Shanmukaraj & Murugan, 2004). It is likely that this factor aids in dictating and limiting the phases accessible for each compound.

**Table 3** Amplitude displacements along *y* of the metal site for $[CH_3NH_3]Co(HCOO)_3$, compound **2** and $[CH_3NH_3]Ni(HCOO)_3$ for different modulation vectors, **q**.

| Compound | **q** | *y* | *T* (K) | **q** | *y* | *T* (K) |
|---|---|---|---|---|---|---|
| $[CH_3NH_3]Co(HCOO)_3$* | 0.1430(2) | 0.0229(4) | 106 | 0.1247(2) | 0.0322(5) | 86 |
| **2** | 0.1429(2) | 0.02033(15) | 70 | 0.1249(2) | 0.0322(2) | 30 |
| $[CH_3NH_3]Ni(HCOO)_3$* | 0.1426(2) | 0.02274(11) | 40 | | | |

*Values for cobalt and nickel compounds obtained from Cañadillas-Delgado *et al*., 2019 and Cañadillas-Delgado *et al*., 2020 references respectively.

**Table 4** Average bond length comparison of the metal environment for $[CH_3NH_3]Co(HCOO)_3$, compound **2** and $[CH_3NH_3]Ni(HCOO)_3$ at 45 K, 30 K and 40 K, respectively.

| Compound | M-O1 (Å) | M-O2 (Å) | M-O3 (Å) |
|---|---|---|---|
| $[CH_3NH_3]Co(HCOO)_3$* | 2.083(3) | 2.101(3) | 2.090(3) |
| **2** | 2.070(2) | 2.087(2) | 2.078(4) |
| $[CH_3NH_3]Ni(HCOO)_3$* | 2.0555(19) | 2.069(2) | 2.059(4) |

*Values for cobalt and nickel compounds obtained from Mazzuca *et al*., 2018 and Cañadillas-Delgado *et al*., 2020 references respectively.

The bulk magnetic properties of **1**, **2** and **3** exhibit a continuous linear trend between the Ni and Co end members. This includes the decrease in the ordering temperature and increase in Curie constant and effective magnetic moment as the cobalt content decreases (Figure 7 and Table S1). Cobalt-nickel solid solutions of molecular frameworks, such as dicyanamides (Lee *et al*., 2016) and hypophosphites, (Marcos *et al*., 1993) report similar trends, with the ordering temperatures increasing almost linearly towards the nickel parent compound. These are rationalised by the strengthening of superexchange interactions as a result of decreasing M-O bond lengths with increasing nickel content (Lee *et al*., 2016). It is plausible that this can explain the magnetic properties observed for these formate compounds as well, where the bond lengths for **2** follow this trend (Table 4). It is possible, however, to observe the strength of the antiferromagnetic correlations in the isothermal magnetisation measurements. The degree of saturation at 5.00(1) T ($M_{5\,T}/M_{sat.}$) decreases from 0.24 for **3**, to 0.17 for **2** and 0.14 for **1**. The decrease in the extent of saturation suggests that by increasing the nickel content, the antiferromagnetic correlations are strengthened.

Both $Ni^{2+}$ and $Co^{2+}$ ions have large orbital contributions to the magnetic moment, due to the second order spin-orbit coupling for $Ni^{2+}$ in an $^3A$ term, and the unquenched orbital moment in $^4T$ term of $Co^{2+}$. As these orbital component do not follow the classical Curie-Weiss dependence on temperature, in all these formate compounds $\theta_{CW}$ varied greatly with the temperature range used to perform the fit.

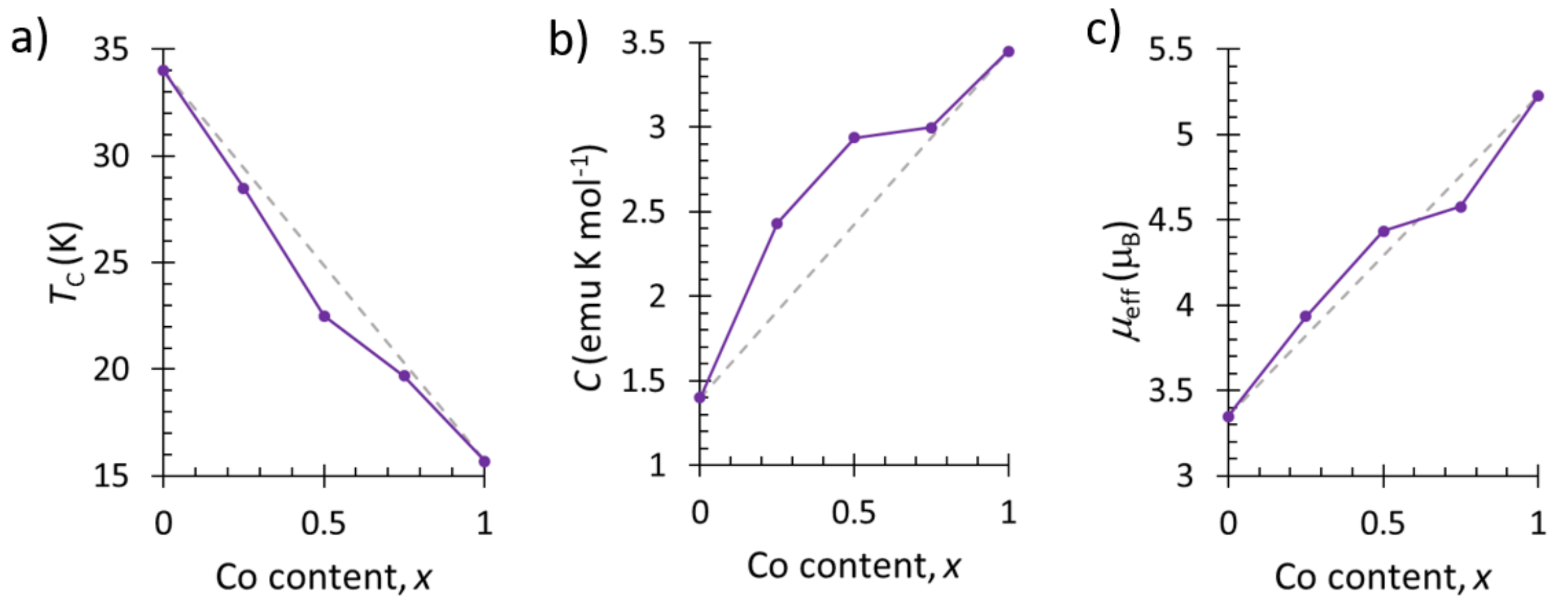

**Figure 7** Summary of magnetometry values obtained from susceptibility measurements of $[CH_3NH_3]Co_xNi_{1-x}(HCOO)_3$ $x$ = 0, 0.25, 0.50, 0.75 and 1. a) The magnetic ordering temperatures, $T_C$. b) Curie constant, $C$ extracted from the Curie-Weiss fit in the renge 150-300 K for the solid solutions and in the ranges 20-300 K and 50-300 K for pure Co and pure Ni compounds, respectively. c) The effective magnetic moment from the Curie-Weiss fit, $\mu_{eff}$. Values for $x$ = 0 and 1 are from their reported values (Pato-Doldán *et al*, 2016; Gómez-Aguirre *et al*., 2016). The dashed lines are guide for the eye between end members.

Compound **2** magnetically orders at $T_C$ = 22.5(7) K in the magnetic superspace group $Pn'ma'(00\gamma)0s0$ with **q** = 0.1249(2)**c***. The moments show weak ferromagnetic ordering, with the uncompensated moment along the $b$ axis. This ordering is broadly comparable to both the Co and Ni end members, which both present weak ferromagnetic superexchange with the nearest neighbours. **2** orders in the same superspace group as the Ni analogue, although with a smaller modulation vector (**q** = 0.1426(2)**c*** for $[CH_3NH_3]Ni(HCOO)_3$ compound). Unlike the magnetic structure of the Ni compound, **2** adopts an improper modulated magnetic structure. It has been reported that by applying a small external magnetic field, approximately 0.05 T, $[CH_3NH_3]Ni(HCOO)_3$ undergoes a transition to an improper incommensurate magnetic phase with collinear moments (Pato-Doldán *et al*., 2023). The activation of any proper magnetic modulations in the structure may be suppressed in **2** by the weaker superexchange pathways resulting from coupling of the Co ions or the modulation of the M-O-C bond angles, which might have the similar effect as the small external magnetic field in the Ni analogue.

#### 4. Conclusion

Three solid solutions of $[CH_3NH_3]Co_xNi_{1-x}(HCOO)_3$, $x$ = 0.25 (**1**), 0.50 (**2**) and 0.75 (**3**), have been synthesised and their nuclear structures and magnetic properties identified through single crystal neutron diffraction and magnetisation measurements. Magnetometry data reveal that their bulk magnetic properties exhibit a linear continuum between the nickel and cobalt end members. The Laue neutron diffraction data permitted a practical method to track the structural behaviour of the compounds and identify the temperature regions of the low temperature modulated phases, with good estimation of the wavevectors.

Monochromatic neutron diffraction data reflect that, similar to the Ni and Co end members, the modulated phases for **2** are induced by the competing hydrogen bond interactions. However, the structural evolution does not follow the same phases as either parent compound. This is likely as a result of the differing Co−O and Ni−O bond lengths which dictate the limits of the atom amplitude displacement modulations. A direct comparison with the pure Co compound reveals that the introduction of a solid solution directly affects the crystalline quality of the sample, as seen by the absence of third-order satellite reflections after the phase transition corresponding to the change of the wavevector from **q** = 0.1429(2)**c*** to **q** = 0.1249(2)**c***.

These solid solutions have shown that through doping of the metal site, the bulk magnetic properties, in particular the magnetic ordering temperature, of $[CH_3NH_3]Co_xNi_{1-x}(HCOO)_3$ can be tuned through the metal ratios. In addition, both the transition temperatures and nature of the nuclear phase transitions can be manipulated via the nickel content. It is worth mentioning that our results advocate that doping the samples increases frustration in the structure, leading to the stabilization of modulated structures over a broader temperature range. Moreover, our findings indicate that the energy barrier separating distinct structural phases is minimal, implying the feasibility of transitioning between them via external stimuli, such as pressure.

The study of modulated structures constitutes an important step in the better understanding of the structure-property relationship of CPs. Despite the sparsity of reported aperiodic molecular frameworks, with understanding of the interactions, this study presents the opportunity to consciously design molecular compounds with the propensity for modulated phases and finer control of their properties.

**Acknowledgements** We acknowledge Dr. Jem Pitcairn for his aid in magnetometry measurements, and ILL for beamtime under proposal numbers 5-41-1157, EASY-779, EASY-846 and EASY-1131. Raw data sets from ILL experiments can be accessed via links provided in references.

# Supporting information

Crystallographic data, in CIF format, for the structures of **2** collected at 70, and 30 K have been deposited at The Bilbao Incommensurate Crystal Structure Database with entry codes Bn1PFN5BD2w and 4YKOEFilc1S, respectively. The crystallographic data, in CIF format, for the commensurate structures at RT of **1**, **2** and **3** can be downloaded from the Cambridge Crystallographic Data Centre through the CCDC reference numbers from 2351889 to 2351891. The magnetic structure of compound **2** at 2 K, in magCIF format, has been deposited at The Magndata Bilbao Structure Database with entry code xxxxxx.

### S1. Experimental

#### S1.1. Synthesis

The synthesis of $[CH_3NH_3]Co_xNi_{1-x}(HCOO)_3$ $x$ = 0.25 (**1**), 0.5 (**2**) and 0.75 (**3**) compounds were followed from reported methods for protonated amine metal formates (Mazzuca *et al*., 2018; Wang, *et al*., 2004). An example (compound **2**) is provided for quantities and timescales used for the synthesis.

Aqueous solutions of $CoCl_2 \cdot 6H_2O$ (119 mg, 0.5 mmol, 1.5 mL), $NiCl_2 \cdot 6H_2O$ (119 mg, 0.5 mmol, 1.5 mL), $CH_3NH_3Cl$ (68 mg, 1 mmol, 3 mL) and NaHCOO (204 mg, 3 mmol, 2 mL) were mixed with *N*-methylformamide ($HCONHCH_3$). The solution was sealed in an autoclave (43 mL) and heated at 413 K for 3 days. The solution was slowly cooled to ambient temperature (approximately 5 h), yielding dark green prismatic crystals (2.3×1.6×1.2 $mm^3$). The crystals were filtered, and dried at room temperature.

Using an analogous method with stoichiometric ratios of $CoCl_2 \cdot 6H_2O$ and $NiCl_2 \cdot 6H_2O$ yield dark green crystals of **1** (2×1.5×1.5 $mm^3$) and dark red crystals of **3** (1.5×1.5×1 $mm^3$).

#### S1.2. Magnetic measurements

Measurements of the magnetic susceptibility were carried out on samples of compounds **1** (12.45 mg), **2** (12.50 mg) and **3** (8.75 mg) using a Quantum Design Magnetic Property Measurements System (MPMS) and a Superconducting Quantum Interference Device (SQUID) magnetometer. Field-cooled (FC) and zero-field-cooled susceptibility measurements were carried out on microcrystalline samples using an applied field of 100 Oe The zero-field-cooled (ZFC) and field-cooled (FC) susceptibility were measured in an applied field of 0.01 T over the temperature range 2 − 300 K. As *M*(*H*) is linear in this field, the small-field approximation for the susceptibility, $\chi(\mathrm{T}) \backsimeq \frac{M}{H}$, where *M* is the magnetisation and *H* is the magnetic field intensity, was taken to be valid. Isothermal magnetisation measurements were carried out at 2 K over the field range −5 to +5 T. Data were corrected for diamagnetism of the sample using Pascal's constants (Bain *et al.*, 2008).

### S1.3. Laue neutron diffraction

The Laue neutron diffraction measurements were collected on the multiple CCD diffractometer CYCLOPS (Cylindrical CCD Laue Octagonal Photo Scintillator, at the ILL, France) which operates with thermal neutrons (Ouladdiaf *et al*., 2011). Single crystals of 2 × 1.5 × 1.5 $mm^3$ $[CH_3NH_3]Co_{0.25}Ni_{0.75}(HCOO)_3$ (**1**), 2.6 × 1.6 × 1.2 $mm^3$ $[CH_3NH_3]Co_{0.50}Ni_{0.50}(HCOO)_3$ (**2**) and 1.5 × 1.5 × 1 $mm^3$ $[CH_3NH_3]Co_{0.75}Ni_{0.25}(HCOO)_3$ (**3**), were mounted on a vanadium pin and placed in a standard orange cryostat. The diffraction patterns were recorded in the temperature range 10 to 120 K, following a ramp of 0.1 K per 30 s. Each Laue diffraction pattern was collected over a period of 15 min with a temperature range of 3 K. The samples were centred on the neutron beam by maximisation of the intensity of several strong reflections in the $x$, $y$ and $z$ directions, after which, a specific orientation was selected and the temperature evolution was collected. Graphical visualisation of the Laue patterns was performed with the ESMERALDA software (Rodríguez-Carvajal *et al*., 2018). Because the neutron beam comprises the wavelength range from 0.8 to 3.2 Å (thermal neutrons) the peaks highlighted in figures S5, S6 and S7 correspond to multiple spots, i.e. there is more than one reflection at different wavelengths contributing to the peak.

### S1.4. Ambient temperature single crystal neutron diffraction (D9)

Hot neutron single crystal diffraction was performed on the four-circle diffractometer D9 (ILL, France) for compounds **1**, **2** and **3**. A monochromatic beam of wavelength $\lambda$ = 0.836 Å was produced using the (220) plane of a Cu crystal in transmission geometry, and a small two-dimensional area detector. Integrated intensities of structural Bragg peaks were collected with standard transverse scans ($\omega$-scans). NOMAD software from the ILL was used for data collection. The structural models were solved using the SUPERFLIP program (Palatinus *et al*., 2007) and refined using Jana2020 (Petříček *et al*., 2023), refining the occupancies of the metal site to obtain the cobalt and nickel ratio of each crystal.

### S1.5. Low temperature single crystal neutron diffraction (D19)

Monochromatic single crystal neutron diffraction data were collected on the four-circle D19 diffractometer (ILL, France) for compound **2**. Neutrons with a wavelength of 1.456 Å were provided by a flat Cu monochromator using the 220 reflection at $2\theta_M$ = 69.91° take-off angle. The sample was placed in a closed-circuit displex cooling device, which was operated following a ramp of 2 K $min^{-1}$. Measurements were taken at 2 K, and at 10 K intervals between 30 and 100 K. NOMAD software from the ILL was used for data collection. Longer data acquisitions were made at 2, 30 and 70 K and were used for the nuclear and magnetic refinements. Unit cell determinations were performed using PFIND and DIRAX programs, and processing of the raw data was applied using RETREAT, RAFD19 and Int3D programs (Duisenberg, 1992; McIntyre *et al*., 1988; Wilkinson *et al*., 1988 & Katcho *et al*., 2021). The data were corrected for the absorption of the low-temperature device using the D19ABSCAN program (Matthewman *et al*, 1982). Nuclear and magnetic models were solved by using the SUPERFLIP program and refined using Jana2020 (Petříček *et al*., 2023). The amplitudes of the first and second order harmonics in the Fourier series of each parameter were refined for all atoms. After the refinement, all those second-order harmonics with values equal to zero were removed in order to avoid overparameterisation. The modulation of the magnetic moments of the metal atoms in the 2 K data set was refined, but the resulting values were close to zero and the statistics did not improve. Consequently, these parameters were removed.

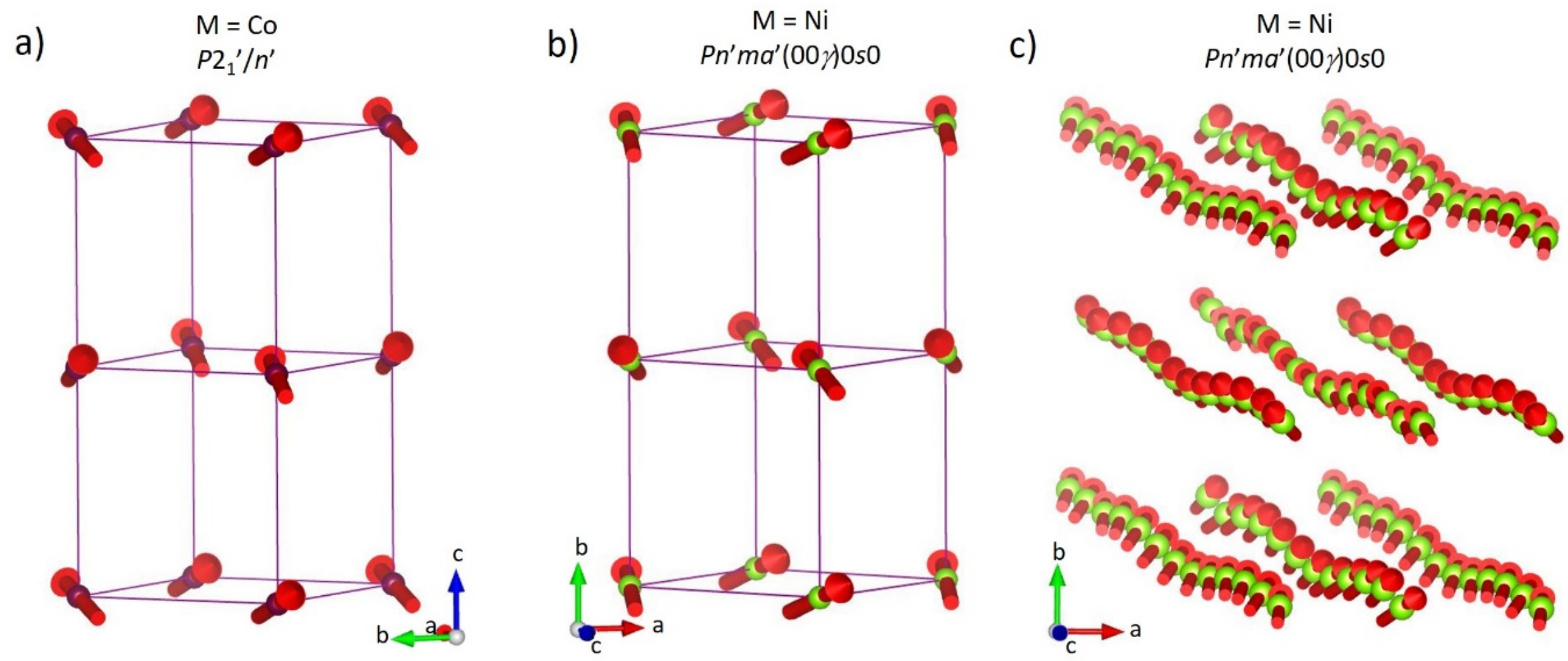

**Figure S1** The magnetic structures for $[CH_3NH_3]M(HCOO)_3$. Both compounds magnetically order with weak ferromagnetic interactions. a) $[CH_3NH_3]Co(HCOO)_3$ in the non-modulated, monoclinic $P2'_1/n'$ magnetic space group, measured at 2 K. b) The average magnetic structure of $[CH_3NH_3]Ni(HCOO)_3$ ordering in the incommensurately modulated *Pn'ma'*(00*γ*)0*s*0 magnetic superspace group, measured at 5 K. The nearest neighbour interactions are displayed with purple lines. c) Modulated magnetic structure of the $[CH_3NH_3]Ni(HCOO)_3$ compound showing the modulation of the atoms, displaced predominately along the *b* axis, and the modulation of the magnetic moments with oscillations in the *ac* plane. The graphical representation has been carried out taking into consideration a super-cell that is ten times of the average structure along the *c* axis in order to include at least a full period.

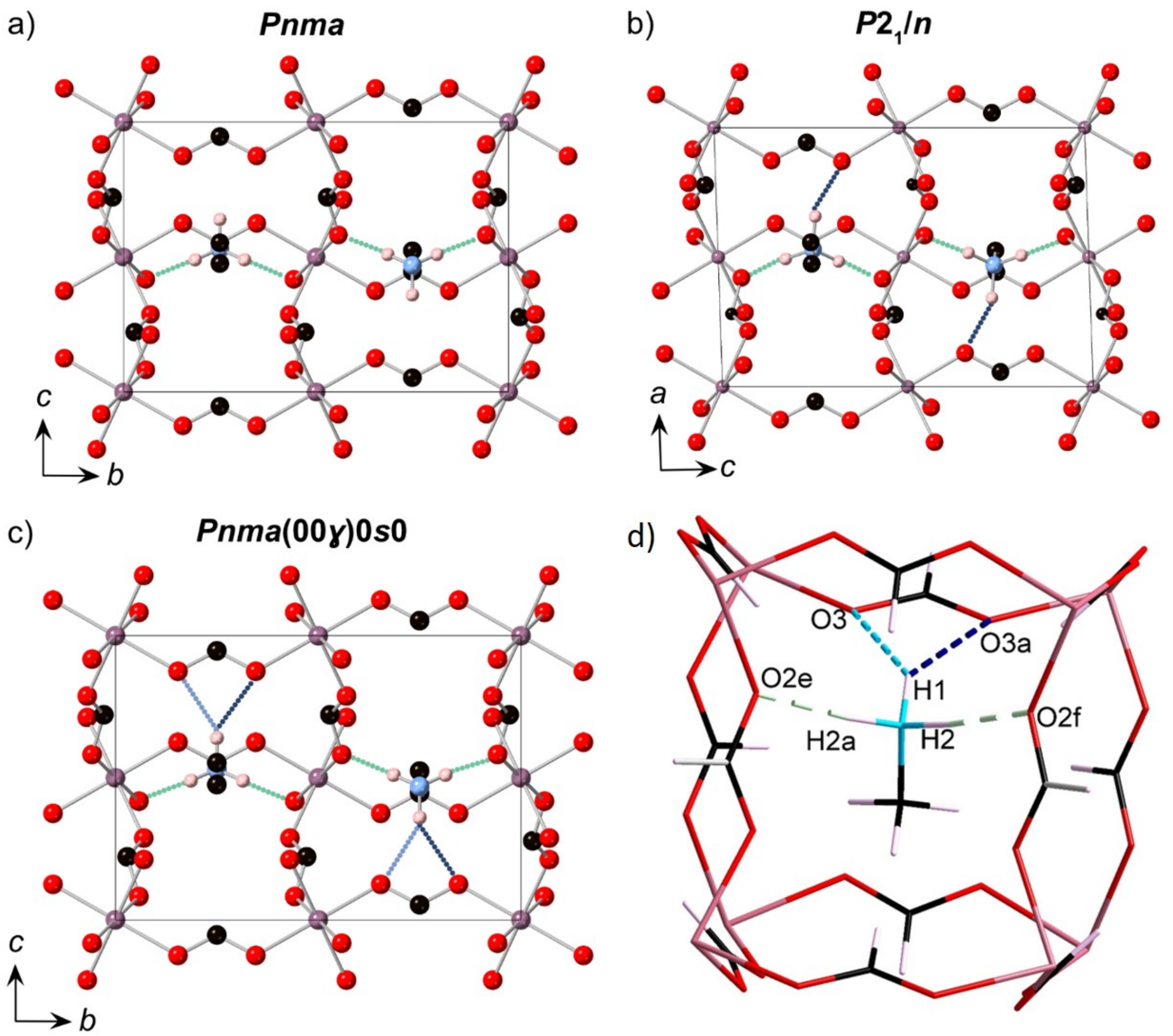

**Figure S2** Schematic view of the possible hydrogen bonds in $[CH_3NH_3]M(HCOO)_3$ M = Co, Ni where green and blue dotted lines indicate the contacts involving H2 and H1 atoms from the $NH_3$ group, respectively. a) For the non-modulated *Pnma* phase at ambient temperature. b) The low temperature, non-modulated $P2_1/n$ phase obtained for M = Co below $T$ = 78 K (Mazzuca *et al*, 2018). c) and d) The average structure of the incommensurately modulated phases *Pnma*(00*γ*)0*s*0, where blue dotted lines highlight the alternated H-bond interactions involving H1 (Cañadillas-Delgado *et al*, 2019). Metal = purple, O = red, C=black, N= blue, H = pale pink.

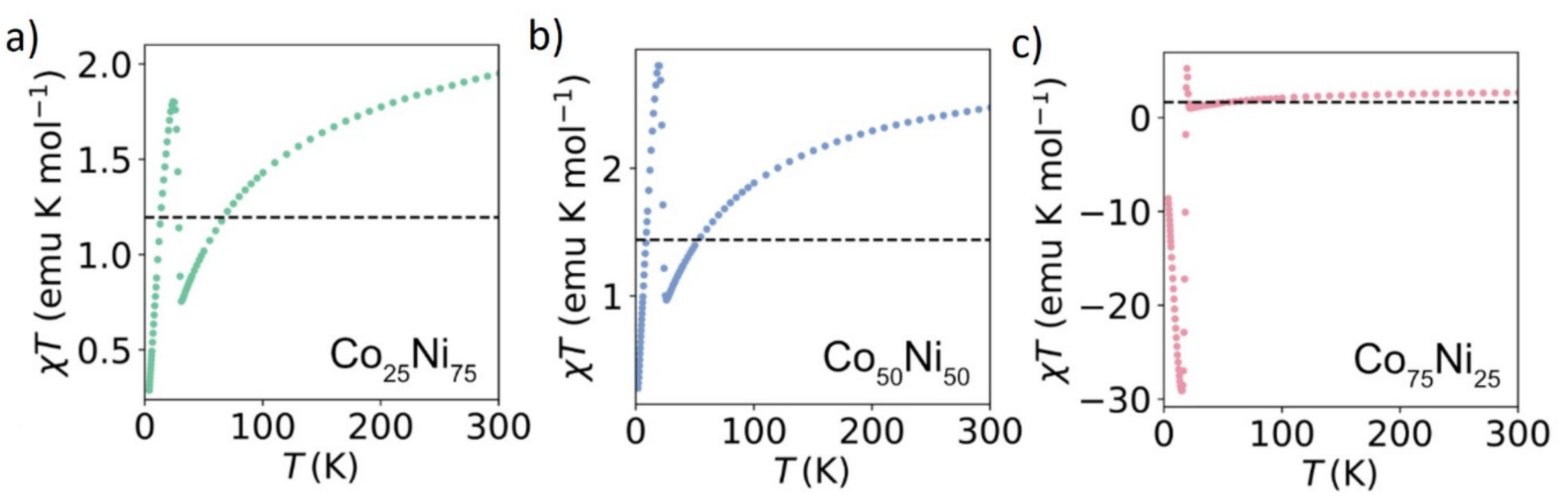

**Figure S3** Plots of the variable temperature magnetic susceptibility product, with the high temperature spin only Curie value, $C_{\text{spin only}}$, indicated with a dashed line. a) Compound **1**, $C_{\text{spin only}}$ = 1.21 emu K mol$^{-1}$, b) Compound **2**, $C_{\text{spin only}}$ = 1.44 emu K mol$^{-1}$ and c) Compound **3**, $C_{\text{spin only}}$ = 1.66 emu K mol$^{-1}$.

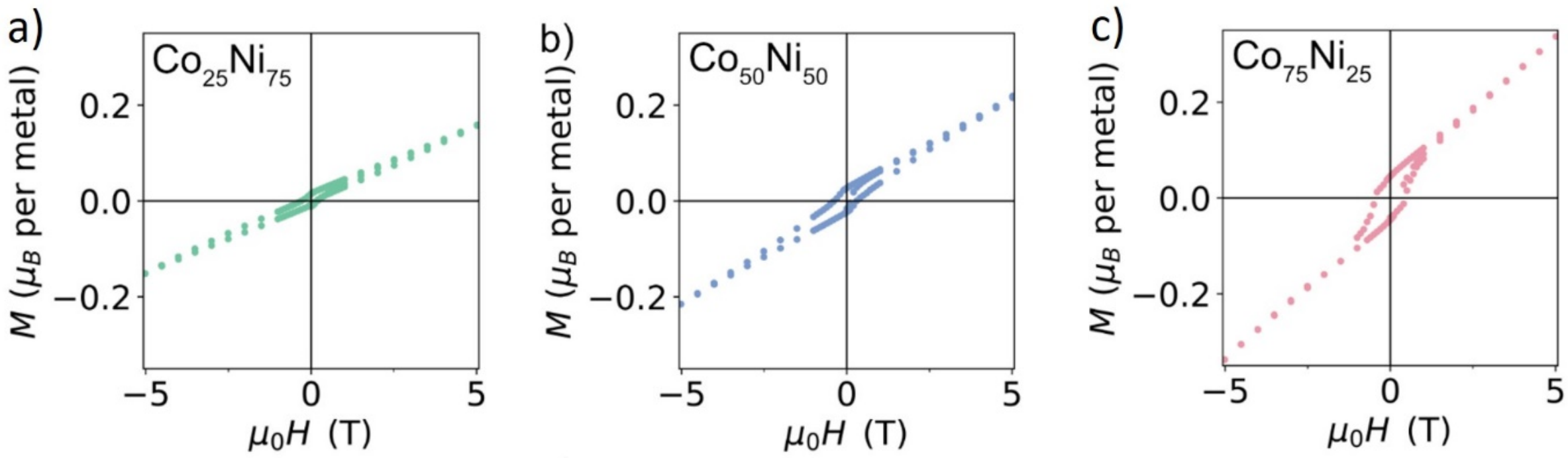

**Figure S4** Isothermal magnetisation measurements for compounds **1** (a), **2** (b) and **3** (c) measured at 2 K between −5.00(1) and 5.00(1) T.

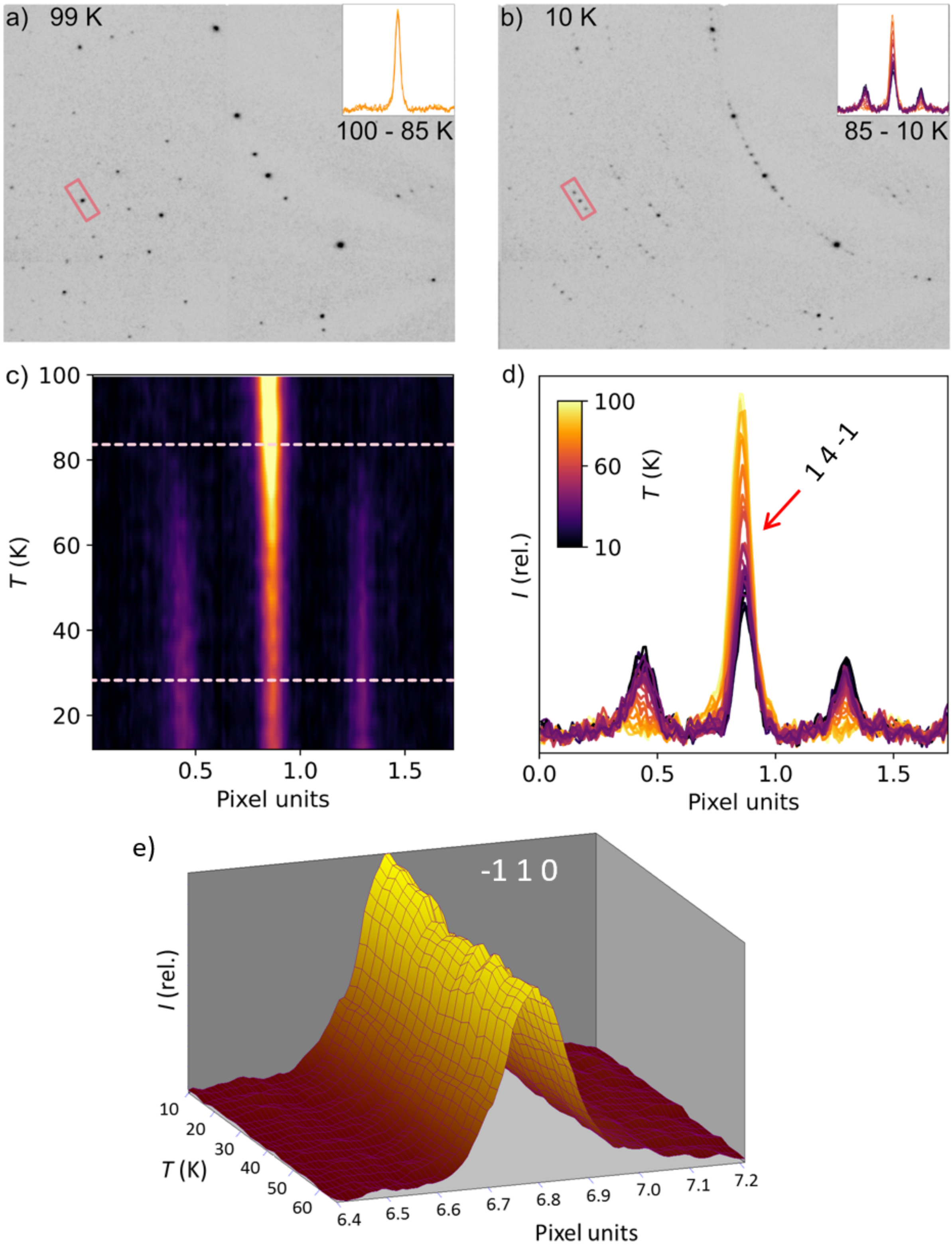

**Figure S5** Laue neutron diffractograms (CYCLOPS, ILL), heating with a ramp of 3 K per image collected during 15 min. for compound **1**. The insets (in a and b) display the integrated intensity for the pixels in the box highlighted in pink, following the multiple spot that contains the 1 4 -1 reflection over the temperature ranges specified. The diffractograms were measured at a) 99 K, showing the non-modulated phase. b) At 10 K showing the modulated phase. c) Temperature evolution in the range 10−100 K for the pixel intensity in the pink box following the multiple spot containing the 1 4 -1 reflection. The pink dashed lines indicate the approximate temperature of the phase transition (structural at 85(3) K, and magnetic, obtained from magnetometry data, at 28.5(5) K. d) Superposition of all the integrated intensities from the insets (a and b) over the temperature range 10−100 K. In c) and d) the multiple spot containing the main reflection 1 4 -1 and the multiple spots corresponding to satellite reflections are observed, on both sides of the main reflection, below about 85(3) K. e) Temperature evolution of the multiple spot containing the -110 reflection, where an increase of intensity at about 28(3) K is observed. The -110 is a structurally forbidden reflection that also increase its intensity because of magnetic order in the pure nickel compound (Cañadillas-Delgado *et al*., 2020).

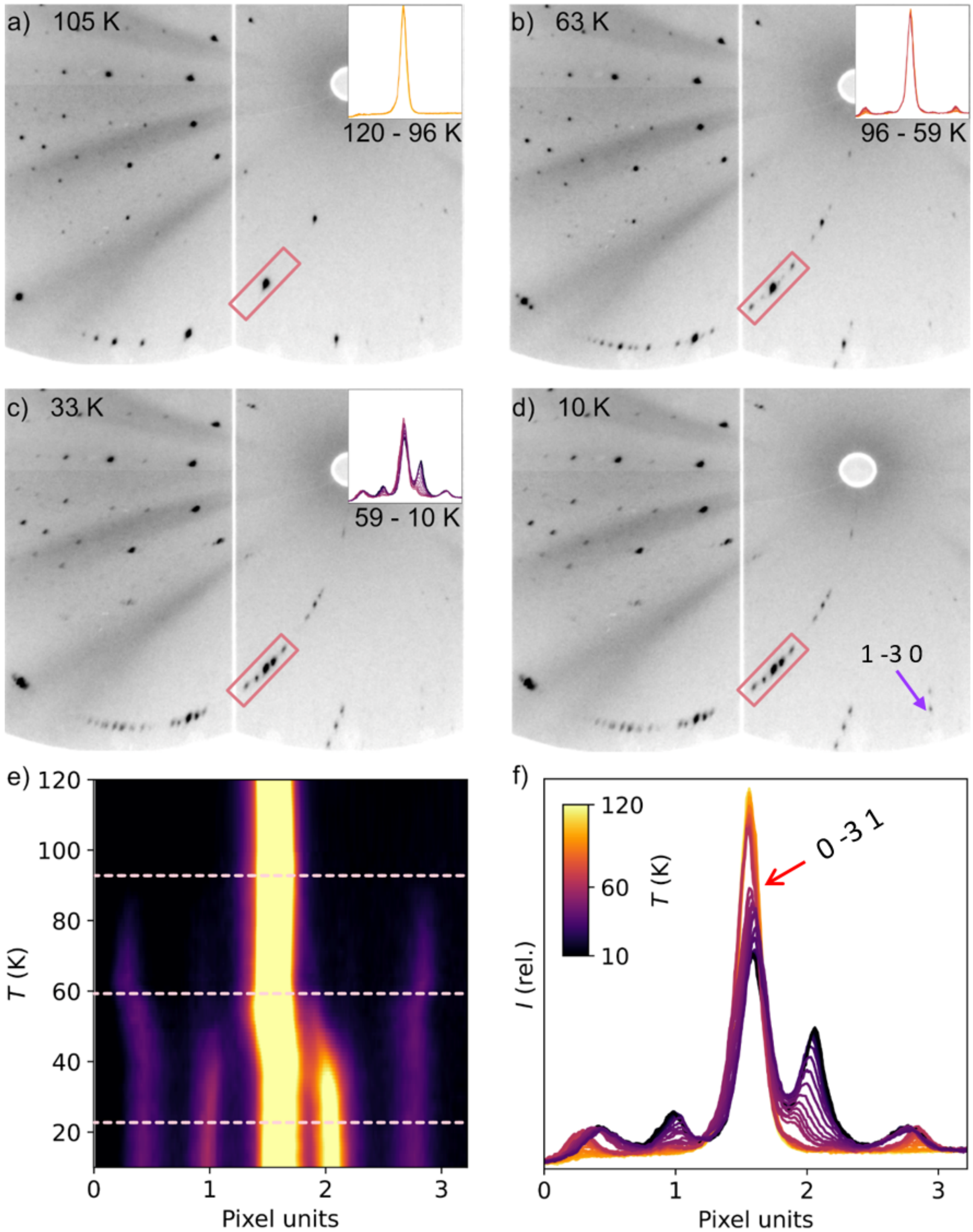

**Figure S6** Laue neutron diffractograms (CYCLOPS, ILL), heating with a ramp of 3 K per image collected during 15 min. for compound **2**. The insets (in a, b and c) display the integrated intensity for the pixels in the box highlighted in pink, following the multiple spot containing the 0 3 -1 reflection over the temperature ranges specified. The diffractograms were measured at a) 105 K, showing the non-modulated phase, b) at 63 K, and c) at 33 K, showing the first and second modulated phases, respectively, and d) at 10 K where a visible magnetic reflection (multiple spot containing the 1 -3 0 reflection) has been highlighted with a purple arrow. e) Temperature evolution over the range 10−120

K for the pixel intensities in the pink box. The pink dashed lines highlight the approximate temperature of the phase transitions (structural at 96(3) K and 59 (3) K and magnetic at 22.5(7) K obtained from magnetometry data). f) Superposition of all the integrated intensities from the insets (a, b and c) over the temperature range 10−120 K. In e) and f) the multiple spot containing the main reflection 0 -3 1 with the multiple spots corresponding to satellite reflections, on both sides of the main reflection, are observed below 96(3) K, and change of distances below 59(3) K.

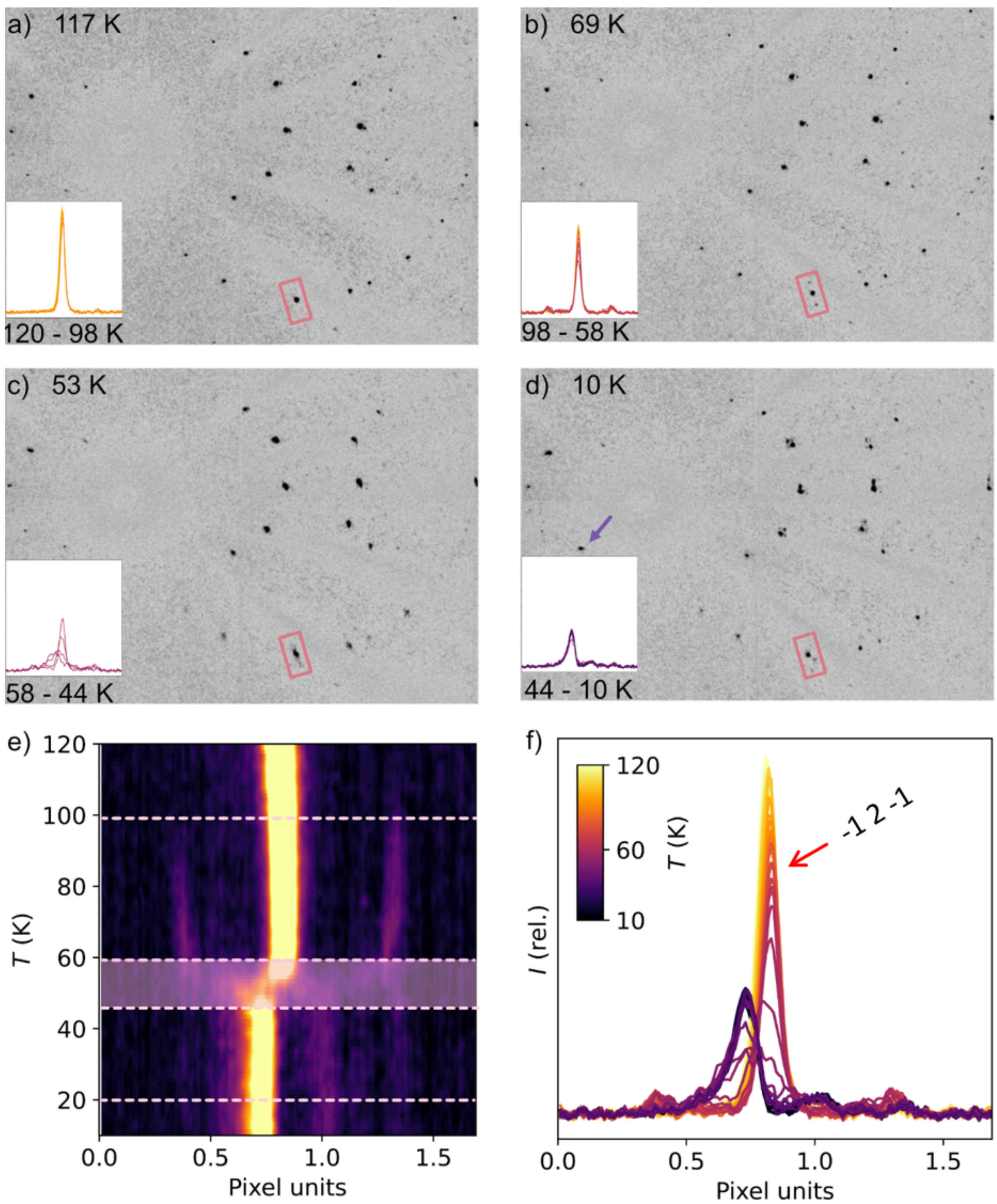

**Figure S7** Laue neutron diffractograms (CYCLOPS, ILL) for compound **3**, following a heating ramp of 3 K per image collected during 15 min. The insets (in a, b, c and d) display the integrated intensity for the pixels in the box highlighted in pink, following the multiple spot containing the -1 2 -1 (*Pnma*) reflection over the temperature ranges specified. The diffractograms were measured at a) 117 K, showing the non-modulated phase, b) at 69 K, showing the modulated phase with weak satellite reflections visible, c) at 53 K, showing the slow phase transition, and d) at 10 K, showing the twinned, non-modulated $P2_1/n$ phase. In d) a magnetic reflection is indicated with a purple arrow. e) Temperature evolution over the 10−120 K range for the pixel intensities in the pink box. The pink dashed lines indicate the approximate temperature of the structural phase transition at 98(3) K, the magnetic order at 19.7(5) K, from the magnetometry data, and the highlighted pink region shows the slow phase transition between 58(3) and 44(3) K. f) Superposition of all the integrated intensities from the insets (a, b, c and d) over the temperature range 10−120 K. In e) and f) the multiple spot containing the main reflection -1 2 -1 with the multiple spots corresponding to satellite reflections, on both sides of the main reflection, are observed between 98(3) K and 44(3) K, with a slight change in the distance between the main and satellite reflections below 58(3) K. Below 44(3) K it could be seen that the pattern changes drastically indicating the appearance of the non-modulated monoclinic twined phase.

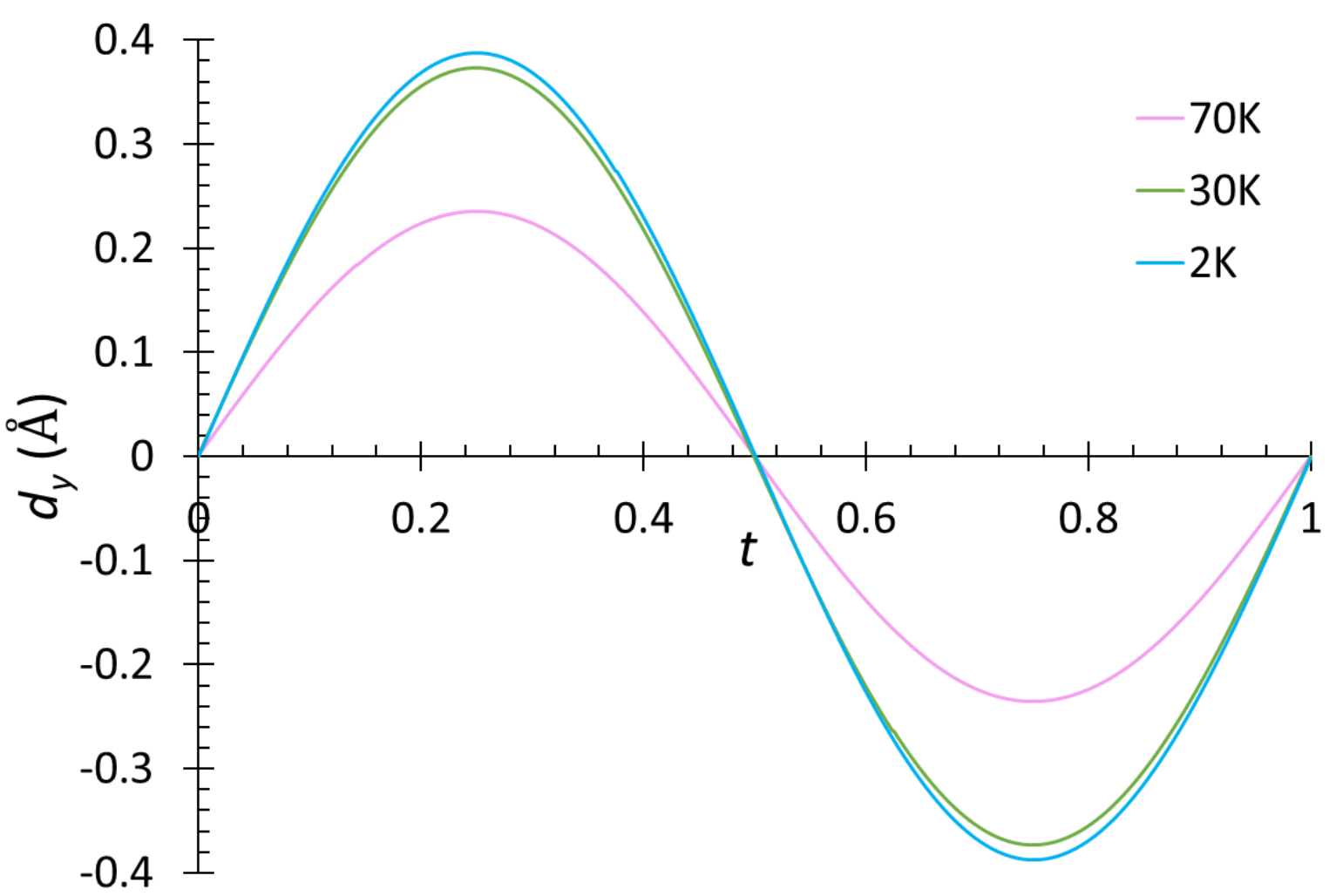

**Figure S8** The modulation function of the metal site exhibiting the displacement along $y$, $d_y$, observed from the refinements of compound **2** measured using the D19 diffractometer (ILL) at 2 K (blue line), 30 K (green line) and 70 K (pink line).

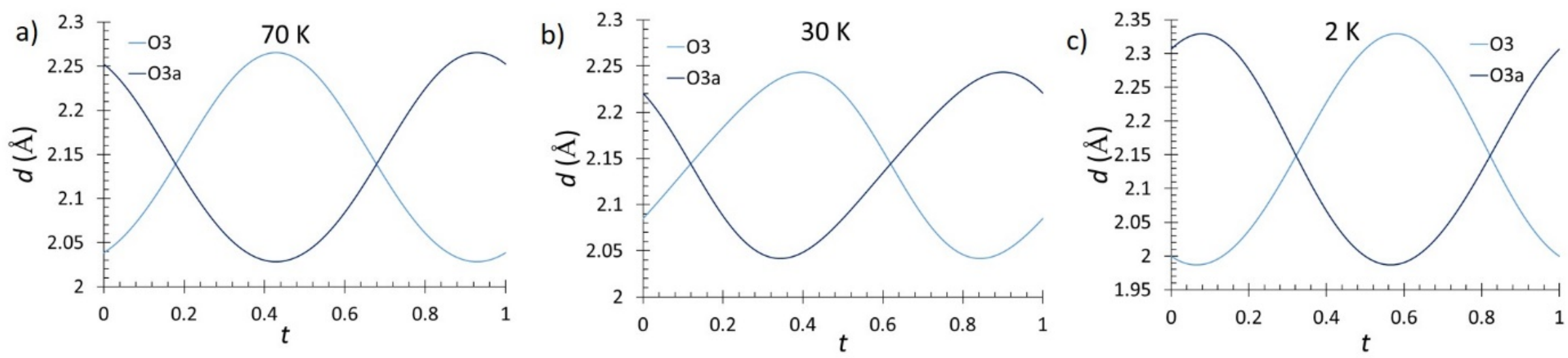

**Figure S9** Modulation of the H1···O3 distances for compound **2** at a) 70 K, b) 30 K and c) 2 K.

**Table S1** Summary of the bulk magnetic properties for $[CH_3NH_3]Co_xNi_{1-x}(HCOO)_3$ $x = 0, 0.25, 0.50, 0.75$ and 1, obtained from magnetisation measurements. $C$ is calculated from the value of $\chi T$ at 300 K. The $\theta_{CW}$ values are calculated from fits over the temperature ranges $150 < T < 300$ K.

| $x$ | 0* | 0.25 | 0.5 | 0.75 | 1** |
|---|---|---|---|---|---|
| S | 1.0 | 1.125 | 1.25 | 1.375 | 1.5 |
| $T_C$ (K) | 34 | 28.5(5) | 22.5(7) | 19.7(5) | 15.9 |
| $\theta_{CW}$ (K) | −64.9 | −70.8(7) | −56.3(1) | −43.6(5) | −43.5 |
| $C$ (cm$^3$ K mol$^{-1}$) | 1.40 | 1.936(4) | 2.461(1) | 2.625(1) | 3.45 |
| $C_{spin\ only}$ (cm$^3$ K mol$^{-1}$) | 1.0 | 1.21 | 1.44 | 1.66 | 1.87 |
| $\mu_{eff.}$ ($\mu_B$) | 3.3 | 3.936(2) | 4.436(2) | 4.578(2) | 5.23 |
| $\mu_{spin\ only}$ ($\mu_B$) | 2.83 | 3.09 | 3.35 | 3.61 | 3.88 |
| $M_{rem.}$ ($\mu_B$ per metal) | | 0.015(1) | 0.027(1) | 0.045(3) | |
| $H_C$ (T) | | 0.15(1) | 0.30(1) | 0.45(1) | |
| $M_{5\,T}/M_{sat.}$ | | 0.140 | 0.175 | 0.245 | |

* Values extracted from Pato-Doldán *et al*., 2016.

** Values extracted from Gómez-Aguirre *et al*., 2016.

**Table S2** Bond lengths distances of the metal environment obtained from the refinement of compound **2** at 70 K, 30 K and 2 K, where it could be appreciated the largest displacement in M-O1 and M-O3 distances at 30 K.

| $T$ (K) | Bond | Average (Å) | Maximum (Å) | Minimum (Å) | Max. displacement (Å) |
|---|---|---|---|---|---|
| 70 | M-O1 | 2.0696(16) | 2.0736(19) | 2.0658(19) | +0.004(2), -0.0038(19) |
| | M-O2c | 2.0835(15) | 2.0879(18) | 2.0793(18) | +0.0044(18), -0.0042(18) |
| | M-O3 | 2.0779(18) | 2.081(2) | 2.075(2) | +0.003(2), -0.003(2) |
| 30 | M-O1 | 2.070(2) | 2.130(3) | 2.014(3) | +0.060(3), -0.056(3) |
| | M-O2c | 2.087(2) | 2.094(3) | 2.081(3) | +0.007(3), -0.006(3) |
| | M-O3 | 2.078(4) | 2.138(5) | 2.009(5) | +0.060(5), -0.069(5) |
| 2 | M-O1 | 2.069(2) | 2.077(3) | 2.061(3) | +0.008(3), -0.008(3) |
| | M-O2c | 2.086(2) | 2.096(3) | 2.077(3) | +0.010(3), -0.009(3) |
| | M-O3 | 2.075(5) | 2.090(5) | 2.066(5) | +0.015(5), -0.009(5) |

*Symmetry code*: c = -*x*+1/2, -*y*+1, *z*+1/2